\documentclass[a4paper,11pt]{article}

\usepackage{jheppub}
\usepackage{amsmath,amsthm,amssymb}
\usepackage{bbm}
\usepackage[utf8]{inputenc}
\usepackage{dynkin-diagrams}
\usepackage{tikz}
\usepackage{tikz-cd}
\usepackage{subcaption}
\usepackage{ytableau}
\usepackage{leftidx}
\usepackage{extpfeil}
\usepackage{scalefnt}

\newcommand{\moy}[1]{\left\langle #1 \right\rangle}

\newcommand{\p}{\partial}
\newcommand{\tr}{\operatorname{tr}}

\newcommand{\Hom}{\operatorname{Hom}}

\newcommand{\im}{\operatorname{Im}}
\newcommand{\Spin}{\operatorname{Spin}}

\newcommand{\vol}{\operatorname{Vol}}
\newcommand{\beq}{\begin{equation}}
\newcommand{\eeq}{\end{equation}}

\newcommand{\Sq}{\operatorname{Sq}}

\newcommand\cG {{\mathcal G}}
\newcommand\cL {{\mathcal L}}
\newcommand\cM {{\mathcal M}}

\newcommand\cP {{\mathcal P}}

\newcommand\cW {{\mathcal W}}
\newcommand\cY {{\mathcal Y}}
\newcommand\fg {{\mathfrak g}}
\newcommand\ft {{\mathfrak t}}

\newcommand\bC {{\mathbb C}}

\newcommand\bR {{\mathbb R}}

\newcommand\bZ {{\mathbb Z}}
\renewcommand\( {\left(}
\renewcommand\) {\right)}
\newcommand\nn {\nonumber}
\newcommand\ds {\displaystyle}
\newcommand\W {{\scriptscriptstyle \mathcal{W}}}
\newcommand\ttb {\mathrm{b}}
\newcommand\rd {\mathrm{d}}
\newcommand\ttf {\mathrm{f}}
\newcommand\ttr {\mathrm{r}}
\newcommand\ttw {\mathrm{w}}
\newcommand\ttcr {\mathrm{cr}}
\newcommand\ttcw {\mathrm{cw}}
\newcommand\tte {\mathrm{e}}
\newcommand\ttm {\mathrm{m}}
\newcommand\ttt {\mathrm{t}}
\newcommand\tts {\mathrm{s}}
\newcommand\ttc {\mathrm{c}}

\newcommand{\nc}{\newcommand}
\nc{\rnc}{\renewcommand}

\rnc{\a}{\alpha}
\rnc{\b}{\beta}
\nc{\g}{\gamma}
\rnc{\d}{\delta}
\nc{\e}{\epsilon}
\nc{\z}{\zeta}
\nc{\f}{\phi}
\nc{\m}{\mu}
\nc{\n}{\nu}
\rnc{\r}{\rho}
\rnc{\k}{\kappa}
\rnc{\l}{\lambda}
\nc{\s}{\sigma}
\rnc{\t}{\tau}
\nc{\w}{\omega}

\nc{\B}{\Beta}
\rnc{\S}{\Sigma}
\rnc{\P}{\Pi}
\rnc{\L}{\Lambda}

\pgfkeys{/Dynkin diagram,edge length=0.7cm,root radius=0.13cm, mark =o}

\title{Line Operators of Gauge Theories on Non-Spin Manifolds}
\author[a,c]{J.P. Ang,}
\author[a]{Konstantinos Roumpedakis,}
\author[b]{Sahand Seifnashri}

\affiliation[a]{C.~N.~Yang Institute for Theoretical Physics\\
Stony Brook University\\
Stony Brook, NY 11794-3840, USA}
\affiliation[b]{Simons Center for Geometry and Physics\\
Stony Brook University\\
Stony Brook, NY 11794-3636, USA}
\affiliation[c]{International Centre for Theoretical Physics\\
Strada Costiera, 11, 34151 Trieste TS, Italy}

\emailAdd{JianPeng.Ang@gmail.com}
\emailAdd{Konstantinos.Roumpedakis@stonybrook.edu}
\emailAdd{Sahand.Seifnashri@stonybrook.edu}

\abstract{We study four-dimensional gauge theories on oriented and non-spin spacetime manifolds. On such manifolds, each line operator arises only either as a boson or a fermion. Based on physical arguments, a method of systematically assigning spin labels to line operators is proposed, and several consistency checks are performed. This is used to classify all possible sets of allowed line operators -- including their spins -- for gauge theories with simple Lie algebras. The Lagrangian descriptions of the theories with these sets of allowed line operators are given. Finally, the one-form symmetries of these theories are studied by coupling to background gauge fields, and their 't Hooft anomalies are computed.}

\preprint{YITP-SB-19-39}

\begin{document}
\maketitle
\section{Introduction}
Line operators in four-dimensional gauge theories are important observables which can be used to classify phases of gauge theories. Wilson lines -- worldlines of heavy electrically charged particles -- obey an area law in a confining phase, while its magnetic analogs, 't Hooft lines -- disorder operators which can be thought of as worldlines of magnetic monopoles -- obey an area law in a Higgs phase \cite{Wilson:1974sk,tHooft:1977nqb}.

While the gauge algebra of a four-dimensional gauge theory fully determines its local operators, it does not fully determine its line operators. Possible line operators are associated to representations of the gauge algebra and its GNO (Langlands) dual \cite{Goddard:1976qe,Kapustin:2005py}, and constraints from locality requires that only a subset of possible operators is allowed. In abelian theories, the charges of line operators have to satisfy the Dirac-Schwinger-Zwanziger quantization condition \cite{Dirac:1948um,Schwinger:1966nj,Zwanziger:1968rs}, and in non-abelian theories they have to satisfy its generalization \cite{Gaiotto:2010be}. Different subsets of allowed line operators correspond to different gauge groups and Lagrangians corresponding to the same gauge algebra.

In addition to their charges, line operators also carry quantum numbers under the spacetime symmetries, transforming under the little group of the Poincar\'e symmetry. On oriented manifolds, this is the group $SO(3)_r$ of spatial rotations, whose projective representations are labeled by the spin, and line operators can be either bosonic or fermionic. However, for non-spin manifolds, there is no spin structure so one cannot define chargeless fermionic lines (worldlines of heavy neutral fermions). Equivalently for a line operator of a given charge, we cannot have both fermionic and bosonic lines, since they could fuse to produce a chargeless fermionic line. However, the existence of charged fermionic lines on such manifolds is allowed and does not require a spin structure. For instance, in the pure $U(1)$ gauge theory the dyon is a fermion~\cite{Jackiw:1976xx,Hasenfratz:1976gr,Goldhaber:1976dp} and the theory does not depend on a spin structure. Therefore on non-spin manifolds, the classification of gauge theories requires the specification of the spin of each allowed line operator, i.e. a spin/charge relation.\footnote{Such spin/charge relations depend on weaker structures rather than a spin structure. For instance, an abelian gauge theory with fermionic Wilson lines requires a $\Spin^\bC$ structure on the spacetime manifold. For the case of $U(1)$ and $SU(2)$ gauge theories this was discussed in \cite{Seiberg:2016rsg,Metlitski:2015yqa,Freed:2016rqq,Cordova:2018acb,Wang:2018qoy, Wan:2019oyr, Hsin:2019gvb,Guo:2017xex,Wan:2018zql,Wan:2019fxh,Wang:2019obe,Wan:2018djl}.} In contrast, on spin manifolds the classification does not depend on the spin, since for any charge there are both bosonic and fermionic lines. This is because the chargeless fermionic lines can be fused with any line in the theory to produce another one with the same charge but different spin. Therefore, on spin manifolds, lines are labeled only by their charges and were classified in~\cite{Aharony:2013hda} for all simple Lie groups.
The purpose of this paper is to classify line operators on oriented but non-spin manifolds, which amounts to consistently specifying the spins of all allowed lines in the theory.

Similarly, on non-orientable spacetime manifolds, we can measure the Kramers properties of Wilson-'t Hooft lines. In this article, we focus on oriented spacetime manifolds and leave the unoriented case for future work. For the case of abelian gauge theories, this classification was previously done in~\cite{Wang:2016cto, Hsin:2019fhf}.

Gauge theories have dynamical particles of spin $1$ -- the gauge bosons -- and hence the spins of line operators are only measurable, at long distances, modulo $1$. Therefore, the theory has line operators labelled by their electromagnetic charges, as well as their spins modulo $1$ -- that is, whether the line operators are bosonic (integer spin) or fermionic (half-integer spin).

Because of the angular momentum of the gauge field, a bound state of two bosonic dyons can be a fermion. For instance, in $U(1)$ gauge theories the bound state of a bosonic electron and a bosonic magnetic monopole of minimal charge is a fermion. More generally, the angular momentum $J_3$, stored in the gauge fields is given by the Dirac pairing between the two dyons
\begin{equation}
2J_3 = \frac{1}{2\pi} (Q_\tte ^1Q_\ttm ^2 - Q_\tte ^2Q_\ttm ^1) \pmod{2\bZ}~,
\end{equation}
where $Q_\tte ^i$ and $Q_\ttm ^i$ are their electric and magnetic charges.

The Dirac quantization condition can be derived by requiring the angular momentum stored in the gauge field to be half-integer, or equivalently the Dirac pairing between any two dyons to be an integer. For a non-abelian gauge theory in the Coulomb phase, it turns out that there is a non-abelian version of the Dirac pairing between line operators which measures the angular momentum stored in the gauge field, see Appendix \ref{app.coulomb}. Motivated by this, we conjecture a way to determine the spin of every allowed line operator, given the spins of two generating lines.

For a non-abelian gauge theory based on Lie algebra $\fg$, the set of line operators are specified by the Weyl orbit of a weight-coweight pairs
\begin{equation}
    w=(\nu,\mu^\vee) \in \Lambda_\ttw \times \Lambda_\ttcw~,
\end{equation}
where $\Lambda_\ttw$ and $\Lambda_\ttcw$ are the weight and coweight lattice of $\fg$ \cite{Kapustin:2005py}. Wilson lines are labeled by the Weyl orbit of a weight $\nu \in \Lambda_\ttw$ which specifies the highest weight of the corresponding representation. Similarly, the 't Hooft lines are labeled by (the Weyl orbit of) a coweight $\mu^\vee \in \Lambda_\ttcw$ which specifies the corresponding GNO charge. The Dirac pairing between two dyonic lines is defined by
\begin{equation}
    \big\langle w, w' \big\rangle_\mathrm{D} := \langle \n, {\m^\vee}'\rangle - \langle \n', \m^\vee \rangle~. \label{introduction.pairing}
\end{equation}
The Dirac quantization condition requires the Dirac pairing to be an integer. Let $s_w$ be the spin of a line ($0$ for boson and $1$ for fermion) whose weights is given by the Weyl orbit of $w \in \Lambda_\ttw \times \Lambda_\ttcw$. We propose that given the spins of lines with weights $w, w' \in \Lambda_\ttw \times \Lambda_\ttcw$, the spin of the line with weight $w+w'$ is
\begin{equation}
\boxed{
        s_{w+w'} = s_w + s_{w'} + \big\langle w, w' \big\rangle_\mathrm{D} \pmod{2\bZ}~.
        } \label{Proposal}
\end{equation}
We further require that the lines whose weights are pure (co)roots to be bosonic, since they correspond to (dual) gluons. In other words, the lines with weights of the form $(\a,0)$ and $(0,\a^\vee)$, where $\a \in \Lambda_\ttr$ and $\a^\vee \in \Lambda_\ttcr$, should be bosonic.

We provide various checks for our proposal. As was showed in \cite{Aharony:2013hda}, for a given group $G$, there are several theories with different generating lines. These theories differ by theta terms, discrete or continuous. The definition of the later on non-spin manifolds, requires a quadratic refinement \cite{Gaiotto:2014kfa}. We show that our proposal is consistent with a quadratic refinement of the discrete theta terms, which we study in detail. These theta terms were determined recently in \cite{Cordova:2019uob}, and our proposal is in complete agreement.

Another check is provided by the Higgsing of a theory with gauge group $G$ down to its maximal torus by adding an adjoint Higgs field. In Appendix \ref{app.coulomb}, we show that after Higgsing, the proposal \eqref{Proposal} correctly produces the spins of line operatros in $U(1)$ theories. Furthermore, in Appendix \ref{App:S and T nonabelian}, we show that the $S$ and $T$ orbtits for $SU(2)$ theories reduce to those of $U(1)$ theories \cite{Wang:2016cto, Hsin:2019fhf}. In Appendix \ref{App: Wu}, we provide an additional cohomological check of our proposal.

The remainder of the paper proceeds as follows. In section \ref{sec:abelian}, we review the classification of abelian charge lattices on non-spin manifolds as a warm up. There are four such theories that we denote by $W_\ttb T_\ttb$, $W_\ttb T_\ttf$, $W_\ttf T_\ttb$ and $W_\ttf T_\ttf$.
The last one is typically referred to as all-fermion electrodynamics and has a gravitational anomaly. We give the UV-completion of all these theories in terms of the Georgi-Glashow model. The Georgi-Glashow model with certain matter whose effective theory is the $W_\ttf T_\ttf$ theory, exhibits the new anomaly discussed in \cite{Wang:2018qoy}. We also give the action of $S$ and $T$ transformations.

In section \ref{sec:rules}, we present our proposal for consistently assigning spin labels to every allowed line operator. For a given gauge group $G$, only a subset of lines satisfying the Dirac quantization condition is allowed, and typically, there are several such sets. For each set, we give the list of requirements which has to be satisfied when assigning spins to lines in the theory. For each set of lines with a given spin, we give a Lagrangian realization and show that our proposal is equivalent to a quadratic refinement of the discrete theta terms.
We further define the action of the $T$-transformation on the weights of the lines, which we use to determine the $SL(2,\bZ)$ orbits.

In section \ref{sec:non-abelian}, we apply our proposal for all simple Lie groups with non-trivial center. We present a complete classification of their charge lattices including the spins of allowed lines. For each case, we give the discrete theta terms and show which theories are related by a $T$ transformation.

In section \ref{sec:background fields}, we consider the couplings to background gauge fields for the one-form symmetry. We review the mixed 't Hooft anomaly in the abelian Maxwell theory. We then consider non-abelian Yang-Mills theories. We show how gauging subgroups of the discrete one-form symmetries yields different theories with the same algebra. This procedure relates all theories in the same family. We further study the requirements for the absence of mixed anomalies for non-abelian theories and we find new anomalies for center gauge groups. 

Finally, in the Appendix we have included additional material to the main discussion of the paper. In Appendix \ref{app:lie.algebra}, we fix our conventions about Lie algebras. In Appendix \ref{app.coulomb}, we perform a consistency check by adding adjoint Higgs fields to the non-abelian theories and flow to abelian theories. In Appendix \ref{App:S and T nonabelian} we give two examples of $S$ and $T$ orbits for the groups $SU(2)$ and $SU(4)$. In Appendix \ref{App: Wu}, we present a cohomological check of or our proposal. In Appendix \ref{App:NormTheta} we fix a normalization for the continuous theta term for any group $G$. Lastly, in Appendix \ref{App:reltheta} we discuss the relation between discrete and continuous theta parameters, in terms of torsion characteristic classes of $G$.

\section{Abelian Theories}
\label{sec:abelian}
As a warm up, we review the classification of $U(1)$ gauge theories on arbitrary oriented  four-dimensional Riemannian spacetime manifolds. These spacetime manifolds are only required to carry an orientation and a metric -- not, for instance, a spin structure. A given theory has a maximal set of line operators that are mutually local, i.e., satisfying the Dirac-Schwinger-Zwanziger (DSZ) quantization condition
\begin{equation}
Q_\tte ^1Q_\ttm ^2 - Q_\tte ^2Q_\ttm ^1 \in 2\pi \mathbb{Z}~,
\end{equation}
where $Q_\tte ^i$ and $Q_\ttm ^i$ are the electric and magnetic charges of two line operators. Apart from these charges, the line operators can be in different representations of the spacetime symmetries, in particular the $(-1)^F$ symmetry. But before studying the properties of the lines under the spacetime symmetries, we first study the electric and magnetic charges of the lines.

\subsection{The Electromagnetic Charge Lattice \label{charge lattice}}
We denote the electric and magnetic charge lattice of the theory as $\Lambda$. In complex coordinate, $\Lambda$ consists of points $z=Q_\tte +iQ_\ttm $, where $Q_\tte $ and $Q_\ttm $ are the electric and magnetic charges of some line operator in the theory. Requiring the DSZ quantization condition, the closeness of the OPE of lines operators, and $\mathcal{C}\mathcal{P}\mathcal{T}$ invariance of the theory, the charge lattice has to satisfy
\begin{align}
\forall z_1, z_2 \in \Lambda &:~ \im\{\bar{z_1}z_2\} = Q_\tte ^1Q_\ttm ^2 - Q_\tte ^2Q_\ttm ^1 \in 2\pi\mathbb{Z}~,\label{DSZ abelian}\\ 
\forall z_1, z_2 \in \Lambda &:~ z_1+z_2 \in \Lambda~,\\
\forall z \in \Lambda &:~ -z \in \Lambda~.
\end{align}
It is easy to see a maximal lattice that satisfy these conditions must be generated by the two closest point to the origin which we denote as $z_1$ and $z_2$, i.e. $\Lambda=\{n_1 z_1+n_2 z_2 | n_1,n_2\in\mathbb{Z}\}$. However, for $U(1)$ theories with a Lagrangian description, there is always the electric excitation of charge $Q_\tte =e$, where $e$ is the coupling constant of the theory. This particle corresponds to the fundamental Wilson line $W(\gamma)=\exp(i\oint_{\gamma} A)$, where $A$ is the $U(1)$ connection. Hence we can set $z_1=e$ and furthermore to satisfy \eqref{DSZ abelian}, we can parametrize $z_2$ such that
\begin{equation}
    z_1 = e\,, \quad z_2 = \tau e\,, \quad \text{where} \quad \tau = \frac{\theta}{2\pi}+\frac{2\pi i}{{e}^2}~, \label{tau}
\end{equation}
and $z_2$ is the charge of the fundamental monopole (dyon), which corresponds to the fundamental 't Hooft line. A priori $\theta$ is a parameter allowed by the DSZ condition and momentarily we will relate it to the Lagrangian description of the theory. Thus all the possible charge lattices are labeled by $\tau$, where $\Lambda_\tau = \{\, (n+m\tau)e \,|\; n,m \in \mathbb{Z} \}$ and it is generated by the fundamental Wilson line $W$ and the fundamental 't Hooft line $T$. The point $(n+m\tau)e\in\Lambda_\tau$ corresponds to the Wilson-'t Hooft line of the form $W^nT^m$, which we denote as $(n,m)$ and its electric and magnetic charges are given by
\begin{equation}
Q_\tte =(n+\frac{\theta}{2\pi}m)e~, \quad Q_\ttm =m\frac{2\pi}{e}~, \label{electromagneticcharges}
\end{equation}
If we ignore discrete spacetime symmetries, the charge lattice $\Lambda_\tau$ classifies all the $U(1)$ theories on spin manifolds, and we could write a local Lagrangian for these theories as
\begin{equation} \label{eqn.maxwellaction}
S_{\tau} =\int_{\mathcal{M}_4} \left(-\frac{1}{2 e^2} f\wedge * f + \frac{\theta}{8 \pi^2} f\wedge f\right)~.
\end{equation} 
where we have identified the charge lattice label $\tau$ with the coupling constant and the theta angle of the theory. Indeed in such a theory, the fundamental monopole has electric charge $\frac{\theta}{2\pi}e$ because of the Witten effect \cite{Witten:1979ey}.

\subsection{Oriented U(1) Gauge Theories}
Now that we have identified the charge lattice, to classify different theories it is enough to label the generating Wilson and 't Hooft lines $W_s$ and $T_s$ with their spins $s$. There are four possibilities: $W_\ttb T_\ttb$, $W_\ttb T_\ttf$, $W_\ttf T_\ttb$, and $W_\ttf T_\ttf$. However the last theory $W_\ttf T_\ttf$, also known as the ``all-fermion electrodynamics", has a gravitational anomaly \cite{Wang:2013zja, Thorngren:2014pza,Kravec:2014aza}. In section \ref{uv completion}, we give UV completions of all these four theories. In the Lagrangian description, the first two theories correspond to $U(1)$ gauge theories and the second two correspond to $\text{Spin}^{\mathbb C}(4):=\text{Spin}(4)\times_{\mathbb Z_2}U(1)$ theories, where we have a $\text{Spin}^{\mathbb C}(4)$ connection instead of a $U(1)$ connection \cite{Metlitski:2015yqa,Seiberg:2016rsg,Hsin:2019fhf}. Hence we get the following theories with their generating lines as
\begin{align}
    U(1)_\ttb &: (1,0)_\ttb\,,\;(0,1)_\ttb\,, &U(1)_\ttf : (1,0)_\ttb\,,\;(0,1)_\ttf~,\\
    {\text{Spin}^{\mathbb C}}_\ttb &: (1,0)_\ttf\,,\;(0,1)_\ttb\,, &{\text{Spin}^{\mathbb C}}_\ttf : (1,0)_\ttf\,,\;(0,1)_\ttf~.
\end{align}
where the labels $\ttb$ and $\ttf$ stand for boson and fermion.

Note that the properties of the Wilson and 't Hooft lines, uniquely determine those of the other lines by fusion, or taking bound states if we think of the lines as world line of heavy classical particles. For instance, the line $(n,m)$ is the bound state of $n$ Wilson lines, and $m$ 't Hooft lines. The quantum numbers of this bound state is the sum of quantum numbers of the individual lines plus the quantum numbers of the electromagnetic field. The electromagnetic field carries a non-trivial quantum number only when there is a minimal pairing between the $(n,0)$ and $(0,m)$ lines \cite{Jackiw:1976xx,Hasenfratz:1976gr,Goldhaber:1976dp}, i.e. $nm \equiv 1 \pmod{2\mathbb{Z}}$. In that case, the non-trivial quantum number is a half-integer angular momentum. For instance, in the $U(1)_\ttb=W_{\ttb}T_\ttb$ theory the dyon $(1,1)$ is a fermion, hence $(1,1)_\ttf$.

\subsection{S and T Transformations} 

In this section we define the action of $S$ and $T$ transformations on the oriented $U(1)$ theories, by defining a map on the line operators of these theories. We define the maps by their action on the $(n,m)_s$ labels denoting the Wilson-'t Hooft line $W^nT^m$ with spin $s$.
Define the action of $S$ and $T$ on the line operators as
\begin{align}
T &: (n,m)_s \mapsto  (n-m,m)_s~,\\
S &: (n,m)_s \mapsto  (m,-n)_s~, \label{abelian s and t}
\end{align}
which leaves the spins invariant. These transformations map the line operators of the oriented $U(1)$ theories to each others. In particular we get the following maps
\begin{equation}
    \begin{array}{ccccccc}
    U(1)_\ttb  & \xmapsto{T} & U(1)_\ttf &\xmapsto{S}  &{\text{Spin}^{\mathbb C}}_\ttb &\xmapsto{T} & {\text{Spin}^{\mathbb C}}_\ttb \\ \hline
    (1,0)_\ttb & \mapsto & (1,0)_\ttb  &\mapsto &(0,-1)_\ttb &\mapsto &(1,-1)_\ttb  \\
    (0,1)_\ttb & \mapsto & (-1,1)_\ttb &\mapsto &(1,1)_\ttb &\mapsto &(0,1)_\ttb  \\
    (1,1)_\ttf & \mapsto & (0,1)_\ttf  &\mapsto &(1,0)_\ttf &\mapsto &(1,0)_\ttf 
    \end{array}
\end{equation}
and the line operators of the ${\text{Spin}^{\mathbb C}}_\ttf$ theory is mapped to itself. However if we also act on the coupling constant $\tau$ as
\begin{equation}\label{S AND T}
	S : \tau \mapsto -\frac{1}{\tau}~, \quad T : \tau \mapsto \tau+1~,
\end{equation}
then $S$ and $T$ become duality transformations (up to gravitational SPT phases) \cite{Witten:1995gf, Metlitski:2015yqa}. In particular, for the $T$-transformation we get a duality between $U(1)_\ttf^{\tau}$ and $U(1)_\ttb^{\tau+1}$, that is shifting $\theta$ by $2\pi$ is equivalent to changing the spin of the monopole\footnote{Another way to see this duality is by noting that from the Witten effect, the $T$-transformation also leaves the electric and magnetic charges invariant. It keeps all the physical properties of the lines invariant, i.e. their charges and their spins.} \cite{Thorngren:2014pza, Benini:2018reh, Hsin:2019fhf, Wang:2018qoy}. Altogether, we get the identifications
\begin{gather}
    U(1)_\ttb^\tau = U(1)_{\ttf}^{\tau+1}, \quad \left({\text{Spin}^{\mathbb C}}\right)_\ttb^{\tau} = \left({\text{Spin}^{\mathbb C}}\right)_\ttb^{\tau+1}, \quad \left({\text{Spin}^{\mathbb C}}\right)_\ttf^{\tau} = \left({\text{Spin}^{\mathbb C}}\right)_\ttf^{\tau+1}~,\\
    U(1)_\ttb^\tau = U(1)_\ttb^{-1/\tau}, \quad U(1)_\ttf^\tau = \left({\text{Spin}^{\mathbb C}}\right)_\ttb^{-1/\tau}, \quad \left({\text{Spin}^{\mathbb C}}\right)_\ttf^{\tau} = \left({\text{Spin}^{\mathbb C}}\right)_\ttf^{-1/\tau}~.
\end{gather}

Thus there are only two classes of physically distinct theories that are not related by continues deformations--the anomalous and the non-anomalous theories--and hence we have the following $\text{SL}(2,\mathbb Z)$ orbits 
\begin{equation}
\begin{tikzcd}
	U(1)_\ttb \arrow[loop left, leftrightarrow,"S"] \arrow[r, leftrightarrow,"T"] & U(1)_\ttf \arrow[r, leftrightarrow,"S"] & {\text{Spin}^{\mathbb C}}_\ttb \arrow[loop right, leftrightarrow,"T"] && {\text{Spin}^{\mathbb C}}_\ttf \arrow[loop right, leftrightarrow,"S\text{, }T"]
\end{tikzcd}
\end{equation}

\subsection{UV Completion of the Oriented U(1) Theories \label{uv completion}}

Here we comment on the Ultra-Violet completions of the oriented $U(1)$ theories we discussed so far. Physically line operators correspond to world line of heavy particles. Therefore to realize microscopic descriptions of these theories, we must provide UV completions where for each line operator there is the corresponding dynamical particle of the same quantum numbers.

The fundamental Wilson line can be UV completed by just adding ordinary matter fields of electric charge $1$. The only non-trivial part is the UV completion of the 't Hooft lines as dynamical monopoles \cite{HOOFT1974276,Polyakov:1974ek}. This can be done by the 't Hooft-Polyakov monopole, which is a classical solution of the Georgi-Glashow model \cite{Georgi:1972cj}; for review see \cite{Harvey:1996ur,Shifman:2012zz}. The Georgi-Glashow model is an $SU(2)$ gauge theory with a scalar Higgs field in the adjoint representation of the gauge group. The 't Hooft-Polyakov monopole is a soliton of finite mass, spin $0$, and unit magnetic charge in the topological sector of the theory. Therefore the $U(1)_\ttb$ and $\left(\text{Spin}^{\mathbb C}\right)_\ttb$ theories can be UV completed by such solutions where there is a dynamical bosonic monopole. Also note that we have the relation $U(1)^{\theta}_\ttf=U(1)^{\theta+2\pi}_\ttb$, therefore to UV complete the $U(1)_\ttf$ theory we just need to add a $\theta^{SU(2)}=\pi$ term in the Georgi-Glashow model as we have the relation
\[
\theta^{U(1)}=2\theta^{SU(2)}~,
\]
with the theta angle in the $U(1)$ theory after Higgsing.

The only remaining theory to discuss is the anomalous $\left(\text{Spin}^{\mathbb C}\right)_\ttf$ theory which was recently studied in \cite{Wang:2018qoy}. The UV completion is the Georgi-Glashow model with an additional Weyl fermion of isospin $\frac32$ which is coupled to the Higgs field via a Yukawa coupling. As it was discussed in \cite{Wang:2018qoy}, the UV theory also has a gravitational anomaly. The additional Weyl fermion gives fermionic zero modes to the 't Hooft-Polyakov monopole solution which makes it a fermion. Hence all of the oriented $U(1)$ theories can be given a microscopic description.

\section{Wilson-’t Hooft Operators and Spin}
\label{sec:rules}
We move on to classify non-abelian gauge theories on oriented spacetime manifolds. In this section, we present and give some justification for our proposal, and give an explicit recipe for constructing the actions of these theories. In the next section, we apply the proposal to each simple Lie group, finding agreement with results in the literature. When the spacetimes are also required to carry spin structures, such a classification was previously carried out in~\cite{Aharony:2013hda}. Here, we extend that discussion to oriented and non-spin manifolds.

As we discussed in the introduction, on non-spin spacetimes the line operators, apart from their electromagnetic charges, are labelled by their spins modulo $1$ -- that is, whether the line operators are bosonic (integer spin) or fermionic (half-integer spin). Mutual locality restricts the set of allowed line operators in a given gauge theory to a maximal set satisfying the Dirac quantization condition. This information can be expressed in terms of a charge lattice decorated with spin labels (boson or fermion).

\subsection{Line Operators}
\label{sec:Line Operators}

We begin by reviewing line operators in gauge theories defined on spin spacetime manifolds, following~\cite{Aharony:2013hda}. Consider a four-dimensional gauge theory with a simple gauge algebra $\fg$. Fix a Cartan decomposition $\ft\subset\fg$ and let $\Lambda_\ttr\subset\Lambda_\ttw\subset\ft^\ast$ be respectively the root and weight lattices, and $\ft^\ast$ the dual Cartan algebra. We denote by $\cW$ the Weyl group. Wilson lines are labelled by representations of $\fg$, which are in bijective correspondence with elements of the Weyl chamber $\Lambda_\ttw/\cW$ -- the highest weights. 't Hooft lines are disorder line operators defined by removing the line from spacetime and imposing boundary conditions on a tubular neighborhood around the line. They are labelled by the dual Weyl chamber $\Lambda_\ttcw / \cW$, where $\Lambda_\ttcw \subset \ft$ is the coweight lattice.\footnote{
The coweight lattice $\Lambda_\ttcw=\Hom(\Lambda_\ttr,\bZ)\subset\ft$ is the dual of the root lattice $\Lambda_\ttr$ (see \eqref{m dot m dual}), or equivalently, the weight lattice of the Langlands dual algebra $^L\fg$.}
More generally, dyonic Wilson-'t Hooft lines are classified by weight-coweight pairs
\begin{equation}
\left[\n,\m^\vee\right]_\W \in ( \Lambda_\ttw \times \Lambda_\ttcw ) / \cW~,
\end{equation}
where $\left[\n,\m^\vee\right]_\W$ is the Weyl orbit of $(\n,\m^\vee)\in \Lambda_\ttw \times \Lambda_\ttcw$. Not all line operators are allowed in a given theory. They are restricted by a non-abelian version of the Dirac-Schwinger-Zwanziger quantization condition~\cite{Kapustin:2005py,Gaiotto:2010be}, which says that the Dirac pairing between any two allowed lines $w=(\nu,\mu^\vee)$ and $w'=(\nu',\mu^\vee{}')$ in a given theory must be integral:
\begin{equation}
\moy{w,w'}_\mathrm{D} := \langle \n, {\m^\vee}' \rangle - \langle \n', \m^\vee \rangle \in \mathbb{Z}~. \label{DSZ non-abelian}
\end{equation}
Here, $\langle \cdot ,\cdot \rangle$ is the pairing induced from the inclusions $\Lambda_\ttw \subset \ft^\ast$ and $\Lambda_\ttcw \subset \ft$ and the natural pairing between $\ft$ and $\ft^\ast=\Hom(\ft,\bR/\bZ)$. This condition is required for the locality of the correlation function between the two line operators, in that the correlation function should remain invariant when the first line operator is transported along a closed surface linking the second line. In the Coulomb phase, this is equivalent to requiring the electromagnetic field between two dyons to have half-integer angular momentum.

The gauge theory contains dynamical adjoint-valued gauge bosons, whose worldlines are labelled by the adjoint representation. Therefore, any line labelled by the sum of an allowed weight with the highest weight of the adjoint representation must itself also be allowed. In other words, it suffices to check the quantization condition \eqref{DSZ non-abelian} on the quotient by the root and coroot lattices
\begin{equation}
\Lambda_\ttw / \Lambda_\ttr \times \Lambda_\ttcw/\Lambda_\ttcr =\hat{Z}(\tilde{G}) \times Z(\tilde{G})~, \label{one-form charges}
\end{equation} 
where $Z(\tilde{G})$ is the center of the simply connected Lie group $\tilde G$ corresponding to $\fg$.\footnote{
$\hat{Z}$ denotes the Pontryagin dual of a group $Z$, $\hat{Z}:=\Hom(Z,\bR/\bZ)$. Note that for a symmetry based on a finite group $Z$, the group elements are inside $Z$, while the charges (the one-dimensional representations of $Z$) are inside $\hat{Z}$.}
A maximal subset of lines obeying the condition~\eqref{DSZ non-abelian}, as well as closure under fusion, corresponds to a subgroup $\cL\subset\hat{Z}(\tilde{G})\times Z(\tilde{G})$ for which the pairing $\moy{w,w'}$ is integral -- in other words, a subgroup which is Lagrangian with respect to the pairing $e^{2\pi i\moy{w,w'}}$~\cite{Aharony:2013hda}.

In this article, we shall consider gauge theories on oriented, but not necessarily spin, spacetime manifolds. As we discussed in the introduction, the main novelty is that without a spin structure, there is a spin/charge\footnote{Here the charge is the gauge charge, which is the weight-coweight pair labeling Wilson-'t Hooft lines, and should not be confused with some global symmetry charge.} relation and a line operator of given (gauge) charge has definite spin (bosonic or fermionic). This is because, if there exist two line operators of same charge (weights) but different spin, one of them could fuse with the $\mathcal{CPT}$ reversal of the other line to produce a neutral fermionic line. But existence of such chargeless fermions (worldlines of gauge singlet fermions) requires a spin structure. To find the possible spin/charge relations, first, we consider the spins of line operators resulting from the fusion of two line operators. Upon fusion, the resulting line generically sits in a reducible representation, and the fusion rules are specified by a complicated operator product expansion (OPE) whose exact form is known only in cases with sufficient supersymmetry~\cite{Kapustin:2006pk,Kapustin:2006hi,Kapustin:2007wm, Drukker:2009id, Gaiotto:2010be, Cordova:2016uwk}. We will not need any details of the OPE -- the only fact used in this article is that the fusion of line operators $[\n,\m^\vee]_\W$ and $[\n',{\m^\vee}']_\W\in \left( \Lambda_\ttw \times \Lambda_\ttcr  \right) / \cW$ includes the line operator $[\n+\n',\m^\vee+{\m^\vee}']_\W$.\footnote{
Note that $[\n+\n',\m^\vee+{\m^\vee}']_\W$ depends on the choice of representatives in the Weyl orbits $[\n,\m^\vee]_\W$ and $[\n',{\m^\vee}']_\W$.}
We propose that the angular momentum stored in the electromagnetic field of the two dyons $w$ and $w'$ is
\begin{equation}
J_3 = \frac{1}{2}\langle w,w'\rangle_\mathrm{D} \pmod{\mathbb Z}~,
\end{equation}
which is integral or half-integral (multiple of $\hbar$) depending on the value of the Dirac pairing
\begin{equation}
    \big\langle w, w' \big\rangle_\mathrm{D} := \langle \n, {\m^\vee}'\rangle - \langle \n', \m^\vee \rangle~ \label{pairing}
\end{equation}
Fusing two dyonic lines with the same spin (boson-boson or fermion-fermion) and odd mutual pairing produces fermionic dyons, while fusing two lines of opposite spins (boson-fermion) and odd pairing produces bosonic dyons. For even paired dyons, the converse holds.

This implies that fusing a line operator with a dynamic gauge boson may change the parity of the pairing \eqref{pairing}, so a line dressed with even or odd numbers of gauge bosons can remain distinguishable at long distances. On the other hand, shifting $\n$ ($\m^\vee$) by twice a root (or coroot) does not change the parity of the pairing and hence preserves spin. Hence, on non-spin manifolds, we propose that the quantization condition \eqref{DSZ non-abelian} needs to be checked on the charges modulo \emph{twice} of the root and coroot lattices. To classify the allowed sets of line operators, and therefore all the possible theories with the same Lie algebra, on a spacetime without spin structure, we consider subsets $\tilde\cL \subset \tilde{\Lambda}$ of the quotient
\begin{equation}
\tilde{\Lambda}:=\Lambda_\ttw / 2\Lambda_\ttr \times \Lambda_\ttcw/ 2\Lambda_\ttcr \label{quotient}
\end{equation}
together with a function $s:\tilde\cL\to\bZ_2$, which specifies the spin of each line operator ($0$ for boson and $1$ for fermion). As line operators are labelled by Weyl orbits, we demand the value of $s$ to be the same on each Weyl orbit of $\tilde\cL$. Furthermore, the data $(\tilde\cL,s)$ should satisfy the following conditions:
\begin{enumerate}
    \item As in~\cite{Aharony:2013hda}, the set of allowed line operators $\tilde\cL\subset\tilde\Lambda$ should be closed under fusion, and be maximal. In other words, it must once again be a subgroup of $\tilde\Lambda$ which is Lagrangian with respect to the pairing $e^{2\pi i\langle w,w'\rangle}$ defined in \eqref{pairing}. Note that Lagrangian subgroups $\tilde\cL\subset\tilde\Lambda$ are in one-to-one correspondence with Lagrangian subgroups $\cL\subset\Lambda\simeq\tilde\Lambda/\((\Lambda_\ttr\times\Lambda_\ttcr)/(2\Lambda_\ttr\times 2\Lambda_\ttcr)\)$. Indeed, roots have integer Dirac pairing with all weights, thus every Lagrangian subgroup $\tilde\cL$ always includes all roots. The main difference between considering $\cL$ and $\tilde\cL$ is that the pairing \eqref{pairing} is defined modulo integers on $\cL$, but is defined modulo even integers on $\tilde\cL$. Moreover, the pairing is odd when there is a half-integer angular momentum contribution from the chromodynamic field to the spin, and it is even when the contribution is integral. 

    \item \label{rule number 2} As discussed above, pure gluons and dual gluons in a gauge theory are bosonic, so we require that $s$ maps elements of the form $(\a,0)$ and $(0,\a^\vee)$ to zero, where $\a \in \Lambda_\ttr/2\Lambda_\ttr$ and $\a^\vee \in \Lambda_\ttcr/2\Lambda_\ttcr$.

    \item The labeling $s$ must be consistent with the fusion rule of two lines taking into account the angular momentum contribution from the chromodynamic field:
    \begin{equation}
        s(w+w') = s(w)+s(w')+ \langle w,w' \rangle_\mathrm{D} \pmod{2}~. \label{spin label}
    \end{equation}
    That is to say, the line $w+w'$ arising from the fusion of two lines with weights $w$ and $w'$ has spin given by the sum of the spins of the individual lines and the contribution from the angular momentum of gauge field, which is $\frac12{\langle w_1,w_2 \rangle}_\mathrm{D} \pmod{\bZ}$.
\end{enumerate}
As a consistency check, we should make sure that the condition \eqref{spin label} is consistent with $s$ being constant on Weyl orbits. Indeed, this follows from the pairing \eqref{pairing} being invariant under a Weyl transformation
\begin{equation}
    \langle w,w' \rangle_\mathrm{D} = \langle x\cdot w,x\cdot w' \rangle_\mathrm{D}~,~~~ x\in\cW~.
\end{equation}

Based on these consistency conditions, in order to classify possible weights and spins $(\tilde\cL,s)$ of gauge theories, it is sufficient to study the set of lines generated by some fundamental Wilson and 't Hooft lines. We denote them respectively as $W(\gamma)$ and $T(\gamma)$ that have supports on a line $\gamma$ in the spacetime. We choose their weights to be some specific weight $\nu_\circ\in\Lambda_\ttw$ and coweight $\mu^\vee_\circ\in\Lambda_\ttcw$, which we define them specifically in a later section. More precisely, we denote the Wilson-'t Hooft lines
\begin{equation}
    W^nT^m(\gamma)~, \quad \quad \text{with weights:} \quad  (n\n_\circ,m\m^\vee_\circ) \in \Lambda_\ttw \times \Lambda_\ttcw~,
\end{equation}
and also label them by the pair of integers $(n,m)$. However, not all of these lines are genuine in a given theory, but to determine the theory it is enough to know the spins of the genuine lines of this form.

\subsection{Lagrangian formulation and discrete theta terms} \label{sec:one-form symmetries}
\subsubsection*{Spin manifolds}
In this section, we review the Lagrangian formulations of the gauge theories considered above, with line operators labelled by $(\tilde\cL,s)$. As a warm up, we begin with spin spacetime manifolds, where the choice of lines is completely determined by $\cL$, which is the projection of $\tilde\cL\subset\tilde\Lambda$ onto $\Lambda=\hat Z(\tilde G)\times Z(\tilde G)$. This will largely be a review of~\cite{Aharony:2013hda}, but will also serve to introduce notation used in the rest of this paper. Let $\Gamma_\ttm$ denote the projection of $\cL$ onto $Z(\tilde G)$. The gauge group of the theory is $G:=\tilde G/\Gamma_\ttm$. Let $\Gamma_\tte:=Z(G)$ denote the center of the gauge group.
The center $Z(\tilde G)$ of the universal covering group and the group of line operators $\cL$ sit in the following two extensions\footnote{The extension \eqref{extension 2} can be viewed as the decomposition of the one-form symmetry group of the $\tilde G$ gauge theory into the subgroup $\Gamma_\ttm$ and the quotient $\Gamma_\tte = Z(\tilde G)/\Gamma_\ttm$. In section \ref{gauging}, we will see that gauging the $\Gamma_\ttm$ symmetry yields the $G=\tilde G / \Gamma_\ttm$ gauge theory with emergent $\hat{\Gamma}_\ttm$ one-form magnetic symmetry.}
\begin{equation} \label{extension 2}
    0\to\Gamma_\ttm \xrightarrow{i} Z(\tilde G)\to\Gamma_\tte\to 0~,
\end{equation}
\begin{equation} \label{extension 1}
    0\to\hat\Gamma_\tte\to \cL \to \Gamma_\ttm\to 0~.
\end{equation}
To see \eqref{extension 1}, note that the line operators with trivial magnetic charge -- Wilson lines -- must be representations of $G$, and therefore sit in $\hat\Gamma_\tte = \hat Z(G) \subset \hat Z(\tilde G)$. In terms of the (co)weight lattices, $\hat\Gamma_\tte = \Lambda_G/\Lambda_\ttr$ and $\Gamma_\ttm = \Lambda_{\ttc G}/\Lambda_\ttcr$ are the electric and magnetic charges of the Wilson and 't Hooft lines respectively for the $G$ gauge theory not coupled to any TQFT, i.e. without any discrete theta term (see figure \ref{chargeslattices.and.symmgroup}).

\begin{figure}
\centering
\[ \begin{array}{|c|c|c|} \hline
\text{1-form} & \text{Charges} & \text{Symmetry group} \\ \hline
\mathrm{Electric} ~&~ \widehat{\Gamma}_\tte= \widehat{Z}(G)=\Lambda_G/\Lambda_\ttr ~&~ \Gamma_\tte = Z(G)=\Lambda_\ttcw/\Lambda_{\ttc G} \\ \hline
\mathrm{Magnetic} ~&~ \Gamma_\ttm = \pi_1(G)=\Lambda_{\ttc G}/\Lambda_\ttcr ~&~ \widehat{\Gamma}_\ttm = \widehat{\pi}_1(G)=\Lambda_\ttw/\Lambda_G \\ \hline
\end{array} \]
\caption{The relation between $\Gamma_\tte,\Gamma_\ttm$, the (co)weight lattices $\Lambda_G,\Lambda_{\ttc G}$, and the one-form symmetries and charges of the plain $G=\tilde{G}/\Gamma_\ttm$ gauge theory. In particular $\Gamma_\tte=Z(G)$ is the electric one-form symmetry group (also known as the center symmetry), and $\Gamma_\ttm=\pi_1(G)$ is the group of magnetic charges carried by the 't Hooft lines, which was denoted $\Gamma$ in~\cite{Gaiotto:2014kfa}.}
\label{chargeslattices.and.symmgroup}
\end{figure}

There are generically many theories with the same gauge group $G$, which differ from each other by discrete theta terms. We now describe these terms. Lagrangian subgroups $\cL\subset\Lambda$ projecting onto $\Gamma_\ttm\subset Z(\tilde G)$ are in one-to-one correspondence with bilinear forms $\eta:\Gamma_\ttm\times\Gamma_\ttm\to\bR/\bZ$ on $\Gamma_\ttm$~\cite{Gaiotto:2014kfa}. This can be seen explicitly as follows: taking \eqref{extension 1} and the dual of \eqref{extension 2}, notice that $\cL$ arises as the pullback
\begin{equation} \label{pullback} \begin{tikzcd}
    0 \ar[r] &\hat\Gamma_\tte \ar[r] \ar[d,equal] &\cL \ar[r] \ar[d,dotted] &\Gamma_\ttm \ar[r] \ar[d,"\eta"] &0 \\
    0 \ar[r] &\hat\Gamma_\tte \ar[r] &\hat Z(\tilde G) \ar[r, "\hat{i}"] &\hat\Gamma_\ttm \ar[r] &0
\end{tikzcd} \end{equation}
with respect to some homomorphism $\eta:\Gamma_\ttm\to\hat\Gamma_\ttm$, or, equivalently, a bilinear form $\eta:\Gamma_\ttm\times\Gamma_\ttm\to\bR/\bZ$. In other words, the maximal subgroup labeling the lines is
\begin{equation}
    \cL = \{(\nu,\mu^\vee)\in\hat Z(\tilde G)\times \Gamma_\ttm:\eta(\mu^\vee)=\hat i(\nu)\} \label{the.lattice}~,
\end{equation}
where $\hat i:\hat Z(\tilde G)\to\hat\Gamma_\ttm$ is the projection map dual to the inclusion $i:\Gamma_\ttm\to Z(\tilde G)$. Physically, this means that the electric charge of a line with magnetic charge $\mu^\vee$ must project onto $\eta(\mu^\vee)$.

The desired discrete theta term is then
\begin{equation}
    2\pi \int_\cM \cP_\sigma(b_\ttm)
    \label{eqn.distheta}
\end{equation}
where $b_\ttm\in H^2(\cM,\Gamma_\ttm)$ is the Brauer class\footnote{
The Brauer class is the characteristic class of the $G=\tilde G/\Gamma_\ttm$ bundle corresponding to the group extension $1\to\Gamma_\ttm\to\tilde G\to G\to 1$. In the physics literature, this is sometimes called the second Stiefel-Whitney class, because the Brauer class coincides with the second Stiefel-Whitney class for the frame bundle, where $G$ is $SO(d)$ and $\tilde G$ is $\Spin(d)$.}
of the gauge bundle, $\sigma:\Gamma_\ttm\to\bR/\bZ$ is a quadratic refinement\footnote{
A quadratic refinement $\sigma:\Gamma_\ttm\to\bR/\bZ$ of a bilinear form $\eta:\Gamma_\ttm \times \Gamma_\ttm\to\bR/\bZ$ is a function satisfying $\sigma(\gamma+\gamma')=\sigma(\gamma)+\sigma(\gamma')+\eta(\gamma,\gamma')$.}
of $\eta$, and $\cP_\sigma:H^2(\cM,\Gamma_\ttm)\to H^4(\cM,\bR/\bZ)$ is the Pontryagin square\footnote{
The Pontryagin square is usually defined as a map $\cP:H^2(\cM,Z)\to H^4(\cM,U(Z))$ where $U(Z)$ is the universal quadratic group of an abelian group $Z$. The universal quadratic group is an abelian group equipped with a quadratic function $\gamma:Z \to U(Z)$, with the property that any quadratic function $q:Z \to A$ taking values in any abelian group $A$ (in this paper, $A=\bR/\bZ$) factors through $\gamma$, i.e. there exists $\tilde{q}:U(Z)\to A$ such that $q = \tilde{q} \circ \gamma$. (See e.g. the Appendix of~\cite{Kapustin:2013qsa} for more details.) Then, we define $\cP_q$ as $\tilde{q}_\ast \cP:H^2(\cM,Z)\to H^4(\cM,A)$.}
operation corresponding to the quadratic form $\sigma$~\cite{Kapustin:2013qsa,Gaiotto:2014kfa}.

On a spin manifold, the discrete theta term \eqref{eqn.distheta} is independent of the choice of quadratic refinement $\sigma$ (and depends only on $\eta$). This gives the complete relation between (1) the set of allowed line operators $\cL$ of the theory, (2) the bilinear form $\eta$, and (3) the discrete theta term \eqref{eqn.distheta} in the Lagrangian formulation. As an example, consider $G=SO(3)$, for which $\tilde G=SU(2)$ and $\Gamma_\ttm=Z(\tilde G)=\bZ_2$. On a spin manifold, there are two distinct choices of line operators, including either the magnetic line $\cL_0=\{(0,0^\vee),(0,1^\vee)\}\subset\hat\bZ_2\times\bZ_2$ or the dyonic line $\cL_1=\{(0,0^\vee),(1,1^\vee)\}\subset\hat\bZ_2\times\bZ_2$. They correspond respectively to the bilinear forms $\eta_0,\eta_1:\bZ_2\times\bZ_2\to\bR/\bZ$ defined by $\eta_0(1^\vee,1^\vee)=0$ and $\eta_1(1^\vee,1^\vee)=\frac{1}{2}$. Each bilinear form has two quadratic refinements, given respectively by
\begin{equation}
    \sigma_0(1^\vee)=0,\quad\sigma_1(1^\vee)=\frac{1}{4},\quad\sigma_0'(1^\vee)=\frac{1}{2},\quad\sigma_1'(1^\vee)=\frac{3}{4}~.
\end{equation}
The corresponding discrete theta terms are, respectively,
\begin{equation}
    \cP_{\sigma_0}=0, \quad \cP_{\s_1}, \quad \cP_{\sigma_0'}=2\cP_{\s_1}, \quad \cP_{\s_1'}=3\cP_{\s_1}~.
    \label{eqn.so3disc}
\end{equation}
Indeed, on spin manifolds $\cP_{\sigma_0'}(b_\ttm)=b_\ttm\cup b_\ttm=b\cup w_2(T\cM)=0 \pmod{2}$, establishing that the discrete theta terms are not dependent on the choice of quadratic refinement.

The discrete theta terms \eqref{eqn.so3disc} may be more familiar to the reader expressed in terms of continuum gauge fields~\cite{Kapustin:2014gua, Gaiotto:2017yup}
\begin{equation}
    \frac{1}{2\pi}\int_\cM f\wedge(\rd a+k\tilde b)+\frac{p(1-k^2)}{2k} k\tilde b\wedge k\tilde b~,
    \label{eqn.contdescr}
\end{equation}
where $a$ is a one-form $U(1)$ gauge field, $\tilde b$ is a two-form gauge field and $f$ is a two-form Lagrange multiplier enforcing the constraint that $\tilde b$ is a $\bZ_k$ gauge field. The extra factor of $(1-k^2)$ is to insure the gauge invariance of \eqref{eqn.contdescr} under
\begin{equation}
b\mapsto b- \lambda,~~~ a\mapsto a+k\, \rd \lambda~,
\end{equation}
which is equivalent to $p(1-k^2)k$ being even.
For even $k$ this factors drops out and $pk$ is indeed even, but for odd $k$ this is necessary to make $pk$ even. Integrating out $f$ and $a$, and making the replacement $k\tilde b/2\pi=b_\ttm$ yields the discrete theta term
\begin{equation}
    2\pi \int_\cM p \cP_{\s_\circ}(b_\ttm) = 2\pi \frac{p(1-k^2)}{2k} \int_\cM b_\ttm \cup b_\ttm + b_\ttm \cup_1 \d b_\ttm~,
\end{equation}
for the quadratic function $\s_\circ(m)=\frac{(1-k^2)}{2k}m^2$. The second term is required since the cup product is not supercommutative at the level of cochains~\cite{Kapustin:2014gua}. In the example of $SO(3)$ above, $k=2$ and $\s_\circ = \s_1$.

\subsubsection*{Non-spin manifolds}
On non-spin manifolds, different quadratic refinements yield distinct discrete theta terms. This is expected, since there is more information contained in the line operators $(\tilde\cL,s)$ of a non-spin gauge theory, which determines the appropriate quadratic refinement, and hence discrete theta term \eqref{eqn.distheta} in the Lagrangian formulation.

Consider first the situation where all purely electric line operators are bosonic, $s(w)=0$ for all $w\in\tilde\cL$ of the form $w=(\nu,0)$. (This corresponds to the gauge group being $G$ rather than $\Spin\text{-}G:=\Spin(4)\times_{\bZ_2}G$.) We claim that the relation between the set $(\tilde\cL,s)$ of allowed line operators and the quadratic function $\sigma:\Gamma_\ttm\to\bR/\bZ$ is given by
\begin{equation}
    \sigma([\mu^\vee]) = \frac{1}{2}\moy{\nu,\mu^\vee}+\frac{1}{2}s(\nu,\mu^\vee),
\end{equation}
where $(\nu,\mu^\vee)\in\tilde\cL$ is the weights of any allowed line, and $[\mu^\vee]\in \Lambda_\ttcw/\Lambda_\ttcr$ denotes the mod $\Lambda_\ttcr$ reduction of the coweight $\mu^\vee\in\Lambda_\ttcw$, similarly $[\nu]\in \Lambda_\ttw/\Lambda_\ttr$ denotes the mod $\Lambda_\ttr$ reduction of the weight $\nu\in\Lambda_\ttcw$. Note that $\moy{\nu,\mu^\vee}$ is defined modulo $2$, so the above definition is single-valued. It can be checked that the right hand side is independent of the choice of representative $\mu^\vee$ as well as choice of line $(\nu,\mu^\vee)$, and furthermore that $\sigma$ is a quadratic refinement of $\eta$, using the conditions we imposed on $(\tilde\cL,s)$ above. We carry out these checks explicitly below.

In the general case, purely electric line operators are not required to be bosonic. In the Lagrangian description, this corresponds to also allowing the gauge group to be $\Spin\text{-}G:=\Spin(4)\times_{\bZ_2} G$, where $\Spin(4)$ is the Lorentz group of the spacetime manifold. In order to account for this, we require $\sigma$ to be extended to a function $\tilde\sigma:\cL\to\bR/\bZ$ defined on all of $\cL$, as follows\footnote{Note that $\tilde\sigma$ is well-defined and only depends on $[\nu]$. That is the RHS of \eqref{eqn.tildesigma} is unchanged if we shift $\nu$ by a root. Suppose $\nu,\nu'\in\Lambda_\ttw/2\Lambda_\ttr$ are two representatives of $[\nu]\in\Lambda_\ttw/\Lambda_\ttr$. Indeed, from the condition \eqref{spin label} we have
\begin{equation}
    s(\nu',\mu^\vee) = s(\nu,\mu^\vee)+s(\nu'-\nu,0)+\moy{\nu-\nu',\mu^\vee} \pmod{2}~.
\end{equation}
Now, $s(\nu'-\nu,0)=0$ according to condition $2$, as different representatives differ by a root. This shows that $\tilde\sigma$ is independent of representative of $[\nu]$. A similar argument can be given for $[\mu^\vee]$.
}
\begin{equation} \label{eqn.tildesigma}
    \tilde\sigma([\nu],[\mu^\vee]) = \frac{1}{2}\moy{\nu,\mu^\vee}+\frac{1}{2}s(\nu,\mu^\vee)~.
\end{equation}
Note that $\tilde\sigma$ is a quadratic refinement of $\eta$ trivially extended to $\cL\times\cL$ as
\begin{equation}
\eta\left(([\nu],[\mu^\vee]),([\nu'],[\mu^\vee{}'])\right) := \moy{\nu,\mu^\vee{}'} = \moy{\nu',\mu^\vee} \pmod{\bZ}~.
\end{equation}
(One can check that $\eta$ defined this way does not actually depend on $\nu$ and $\nu'$, and coincides with the $\eta$ defined above corresponding to $\cL$.) Indeed, this follows immediately from the conditions \eqref{spin label} since
\begin{equation}
\tilde\sigma([\nu+\nu'],[\mu^\vee+\mu^\vee{}']) - \tilde\sigma([\nu],[\mu^\vee]) - \tilde\sigma([\nu'],[\mu^\vee{}']) = \eta\left(([\nu],[\mu^\vee]),([\nu'],[\mu^\vee{}'])\right)~.
\end{equation}

The discrete theta term corresponding to the generalized quadratic refinement \eqref{eqn.tildesigma} is
\begin{equation} \label{eqn.disctheta2}
2\pi \int_\cM \cP_{\tilde\sigma}(b)~,
\end{equation}
where $b\in H^2(\cM,\cL)$ is a dynamical symmetry current. $b$ can be related to the Brauer class $b_\ttm\in H^2(\cM,\Gamma_\ttm)$ mentioned above, as well as another class $b_\tte\in C^2(\cM,\hat\Gamma_\tte)$ via the decomposition
\beq b = \tilde b_\ttm + b_\tte~, \eeq
where $\tilde b_\ttm\in C^2(\cM,\cL)$ is some lift of $b_\ttm$ to $\cL$. The current $b_\tte$ satisfyies a twisted closure condition
\beq \delta b_\tte = b_\ttm^\ast e~, \eeq
where $e\in H^3(B^2\Gamma_\ttm,\hat\Gamma_\tte)$ is the extension class of $\cL$. In practice, the discrete theta term will only involve a $\bZ_2$ subgroup of $\hat\Gamma_\tte$, and we only need deal with a mod $2$ reduction of $b_\tte$.\footnote{
Fermionic electric lines are only possible when $\Gamma_\tte=Z(G)$ has a $\bZ_2$ subgroup, which couples to $(-1)^F\in\Spin(4)$ of the Lorentz symmetry.}
$b_\tte$ is only turned on when there are fermionic pure electric lines, identifying the $\bZ_2$ subgroup of the gauge group with $(-1)^F$ and imposing a spin/charge relation on the theory.

In the simple case where $\cL$ is a direct product $\hat\Gamma_\tte\times\Gamma_\ttm$, $b_\tte$ is closed, and the discrete theta term can be written explicitly in terms of the conserved currents $b_\tte$ and $b_\ttm$ as
\begin{equation}
    2\pi \int_\cM \cP_{\tilde\sigma}(b_\tte,b_\ttm) = 2\pi \int_\cM \frac{s'}{2}b_\tte\cup b_\tte + \cP_\sigma(b_\ttm)~,
\end{equation}
where $s'$ is the spin of any pure electric line generating $\Gamma_\tte$, and $\sigma$ is $\tilde\sigma$ restricted to $\{0\}\times\Gamma_\ttm$. By integrating out $b_\tte$, it can be explicitly seen that this coupling turns the gauge group from $G$ to $\Spin\text{-}G=\Spin(4)\times_{\bZ_2}G$ when $s'=1$.

As an example, take again the example of $G=SO(3)$, focussing on the case where the line operator is magnetic, which we denoted by $\cL_0$ above. Corresponding to $\cL_0$, there are two possible assignments of spins $s$ on $\tilde\cL_0$. Identifying $\Lambda_\ttcw/2\Lambda_\ttcr\simeq\bZ_4$, we denote the lines as
\begin{equation}
    \tilde\cL_0 = \{(0,0^\vee),(0,1^\vee),(0,2^\vee),(0,3^\vee),(2,0^\vee),(2,1^\vee),(2,2^\vee),(2,3^\vee)\}.
\end{equation}
The two possible assignments are $s^{(1)}$ assigning $(2,1^\vee)$ and $(2,3^\vee)$ to be fermionic and all other lines to be bosonic, or $s^{(2)}$ assigning $(0,1^\vee)$ and $(0,3^\vee)$ to be fermionic, and all other lines bosonic. These theories are denoted as $SO(3)_{0,\ttb}$ and $SO(3)_{0,\ttf}$ in the discussion around \eqref{lines of SO(3)} and correspond respectively to the two quadratic refinements
\begin{equation}
    \sigma^{(1)}(1^\vee) = 0\,,\quad\quad \sigma^{(2)}(1^\vee) = \frac{1}{2}~,
\end{equation}
of the bilinear form $\eta=\eta_0(1^\vee,1^\vee)=0$.  Therefore, the corresponding discrete theta terms are
\begin{equation}
    0\,,\quad\quad 2\pi \int_\cM \frac{1}{2}b_\ttm \cup b_\ttm = 2\pi \int_\cM \frac{1}{2} b_\ttm\cup w_2(T\cM)~,
\end{equation}
respectively, recovering the result that the discrete theta term $\pi \, b_\ttm\cup w_2(T\cM)$ changes the spin of the magnetic lines.

\subsection{\texorpdfstring{$T$ transformations}{T transformations}} \label{sec:T transformation}
Some theories with the same gauge group but different discrete theta terms can be related to one another by shifts of the usual continuous theta angle\footnote{
See Appendix \ref{App:NormTheta} for conventions for the normalization of theta terms.}
\begin{equation}
    \frac{\theta}{2} \int_\cM \tr{\frac{f}{2\pi}\wedge \frac{f}{2\pi}}~. \label{theta term}
\end{equation}
A discrete theta term \eqref{eqn.distheta} changes the set $(\tilde\cL,s)$ of line operators of a theory \cite{Aharony:2013hda}, while the continuous theta term \eqref{theta term} does not change the charge lattice. Nevertheless, in some situations, instantons fractionalize and the periodicity of $\theta$ is increased.\footnote{
This heralds an anomaly between the $(-1)$-form symmetry $\theta\mapsto\theta+2\pi$ and the $1$-form electric center symmetry of the $\tilde G$ gauge theory~\cite{Aharony:2013hda,Gaiotto:2014kfa}. It can also be understood as an anomaly in the space of coupling constants~\cite{Cordova:2019jnf,Cordova:2019uob}.}
When this happens, shifting $\theta$ by $2\pi$ becomes equivalent to adding a discrete theta term.\footnote{
See Appendix \ref{App:reltheta} for a mathematical discussion of this phenomenon.}
More precisely, a theory (with a given discrete theta term) at $\theta=\theta_0$ is dual to another theory (with a different discrete theta term) at $\theta=\theta_0+2\pi$. We refer to the duality map between the two theories as the $T$ transformation
\begin{equation}
    T: G^\theta_p \to G^{\theta+2\pi}_{p-\Delta p}~, \label{T-duality}
\end{equation}
where $G^\theta_p$ denotes the theory with gauge group $G$ with parameter $\theta$ and discrete theta parameter labelled by $p$, as in~\cite{Aharony:2013hda}.

\subsubsection*{Relation between continuous and discrete theta terms}
In the following, we compute the exact discrete theta term, schematically denoted $\Delta p$ in \eqref{T-duality}, generated by increasing $\theta$ by $2\pi$.

First, we choose a special basis for the weights and coweights, as follows. Assume for now that $\fg$ is a simple Lie algebra, but \emph{not} of type D$_N$ with $N$ even, so that the center $\Lambda_\ttcw/\Lambda_\ttcr=Z(\tilde G)$ is a cyclic group $\bZ_M$. Choose a coweight $\mu_\circ^\vee\in\Lambda_\ttcw$ whose class $[\mu_\circ^\vee]$ in $Z(\tilde G)$ generates it. Now, choose a weight $\nu_\circ\in\Lambda_\ttw$ such that its class in $\hat Z(\tilde G)$ has weight one with respect to the chosen coweight; that is,
\begin{equation}
    \moy{\nu_\circ,\mu_\circ^\vee} = \frac{1}{M}~. \label{WT lines basis}
\end{equation}
Every weight (coweight) can be expressed as an integer multiple of $\nu_\circ$ ($\mu_\circ^\vee$) modulo roots (coroots), i.e. $[\nu]\equiv n[\nu_\circ]\pmod{\Lambda_\ttr}$ and $[\mu^\vee]\equiv m[\mu_\circ^\vee] \pmod{\Lambda_\ttcr}$, for some $n,m\in\bZ_M$.\footnote{
$n$ and $m$ can be taken modulo $M$ since $M$ times of $\nu_\circ$ ($\mu_\circ^\vee$) is a root (coroot).}
For example, for $\fg=\mathfrak{su}(N)$ the center is isomorphic to $\bZ_N$ and we choose\footnote{Such a choice of generator for $Z(\tilde{G})$, is equivalent to choosing a ring structure for $Z(\tilde{G})$ whose unit element under multiplication is $[\mu_\circ^\vee]$. Writing cup product in the action requires such a ring structure, but here such a choice is made in \eqref{sigma0} which defines a pairing on $H^2(\cM,Z(\tilde{G}))$.}
$\mu^\vee_\circ:=\mu^\vee_1$ corresponding to the fundamental representation of the dual algebra $^L\fg$, and $\nu_\circ:=\nu_{N-1}$ to be the highest weight of the anti-fundamental representation of $\fg$ which acts with charge $1/N$ on the defining $N$-dimensional representation.

We call a line of weight $(\nu_\circ,0)$ a basic Wilson line $W$, and a line of weight $(0,\mu_\circ^\vee)$ a basic 't Hooft line $T$. Not all possible lines are allowed in a given theory. Consider, for example, $\tilde G$ gauge theory. $W$ is an allowed line operator, while $T$ is not. Rather, $T$ is the boundary of the (open) Gukov-Witten surface operator $U_\tte(\Sigma)$ implementing the $Z(\tilde G)$ electric one-form symmetry. Then, \eqref{WT lines basis} leads to the following relation
\begin{equation}
    U_\tte(\Sigma)W(\gamma) = e^{2\pi i\moy{\nu_\circ,\mu_\circ^\vee} \operatorname{link}(\Sigma,\gamma)} W(\gamma) = e^{2\pi i/k \operatorname{link}(\Sigma,\gamma)} W(\gamma)~.
\end{equation}
Thus, \eqref{WT lines basis} is equivalent to choosing a basis such that the fundamental line $W$ has charge one under the fundamental electric symmetry generator. A similar argument can be made for $\tilde G/Z(\tilde G)$ theory where $T$ is now an allowed line operator and $W$ is the boundary of the fundamental magnetic surface operator implementing the $\widehat Z(\tilde G)$ magnetic one-form symmetry which can be written as
\begin{equation}
    U_\ttm(\Sigma) = \exp{\left(2\pi i [\nu_\circ] \oint_{\Sigma} b_\ttm \right)}~,
\end{equation}
where
\footnote{\eqref{WT lines basis} implies that $[\nu_\circ]$ generates the magnetic symmetry group $\widehat{Z}(\tilde{G})$.}
$b_\ttm \in H^2(\cM,Z(\tilde{G}))$ is the Brauer class of the gauge bundle.

Next, consider the effect of adding discrete theta terms on the line operators of a theory with gauge group $\tilde G/\bZ_k$, where $\bZ_k$ is some subgroup of $Z(\tilde G)=\bZ_M$, so $M=kk'$ for some integer $k'$. Quadratic functions on $\Gamma_\ttm=\bZ_k$ can be written as $\sigma=p\, \sigma_\circ:\bZ_k\to \bR/\bZ$ for some $p$, where, following the conventions of~\cite{Gaiotto:2014kfa}, we define
\footnote{$\sigma_\circ$ is so chosen because it is a quadratic refinement of the bilinear form $\eta_\circ(m [\mu_\circ^\vee],m' [\mu_\circ^\vee])=mm'/k$. For $k$ odd, this is the unique refinement, reflecting the fact that the basic 't Hooft line must be bosonic, while for $k$ even, there are two refinements, $\sigma_\circ$ and $(1+k)\sigma_\circ$, since the basic 't Hooft line can be bosonic or fermionic.
}
\begin{equation}
    \sigma_\circ(m k'[\mu_\circ^\vee]) = \begin{cases} \ds\frac{m^2}{2k} ~~~& k\text{ even}~, \\ \ds\frac{(1-k)m^2}{2k} ~~~& k\text{ odd}~. \end{cases} \label{sigma0}
\end{equation}
(Note that $k'[\mu_\circ^\vee]$, with $[\mu_\circ^\vee]$ as defined above, generates the $\bZ_k$ subgroup of $\bZ_M$.)
In this paper, we often identify $\bZ_k$ with integers modulo $k$, and $k'[\mu_\circ^\vee]$ with $1\in \bZ/k\bZ$, so in both cases this can be written as
\begin{equation} \label{refinement}
    \sigma_\circ(m)= \frac{1-k^2}{2k}m^2~.
\end{equation}
According to our proposal \eqref{eqn.tildesigma}, the theory defined by adding the discrete theta term
\begin{equation} \label{B cup B}
    2\pi p \int_\cM \cP_{\sigma_\circ}(b_\ttm)
\end{equation}
has the bosonic line labelled by $(p(1-k^2)\nu_\circ,k'\mu_\circ^\vee)_\ttb$ as a genuine line operator.

Meanwhile, a shift $\theta\mapsto\theta+2\pi$ of the continuous theta angle \eqref{theta term} results in the following shift in the line operators~\cite{Kapustin:2005py}
\begin{equation}
    T: (\n,\m^\vee)_s \mapsto (\n-{\m^\vee}^\ast,\m^\vee)_s~, \label{T transformation}
\end{equation}
where $^\ast$ is a map from $\Lambda_\ttcw$ to $\Lambda_\ttw$ induced by the Killing form and the inclusions $\Lambda_\ttcw\subset\ft$ and $\Lambda_\ttw\subset\ft^\ast$, and $s$ is the spin label. The Killing form, denoted by $\tr$ in \eqref{theta term}, is normalized such that a generator of $\Lambda_\ttcw$ is mapped to a generator of $\Lambda_\ttw$. More explicitly, the normalization is such that a simple coroot $\alpha_i^\vee$ is mapped to a simple root with the following prefactor~\cite{Kapustin:2005py}
\begin{equation} \label{roots.coroots}
    {\a_i^\vee}^\ast=\frac{\tr\a^\vee_i\a^\vee_i}{2} \a_i~,
\end{equation}
where the length-squared of short coroot(s) is $2$; see Appendix \ref{app:lie.algebra} for our Lie algebra conventions. The $2\pi$ shift maps one allowed set $(\tilde\cL,s)$ of line operators to another allowed set, and therefore maps between theories with the same gauge group.~\footnote{
$T$ transformations do not affect the spins of purely electric lines (or equivalently, the gauge group), and therefore do not affect the $b_\tte$ dependence of discrete theta terms \eqref{eqn.disctheta2}.}

We now derive the relation between continuous and discrete theta terms. $T$ sends a line $(\nu,k'\mu_\circ^\vee)$ with fundamental magnetic charge to $(\nu-k'\mu^\vee_\circ{}^\ast,k'\mu_\circ^\vee)$. According to \eqref{eqn.tildesigma}, the shift in its electric charge is equivalent to adding a discrete theta term \eqref{eqn.distheta} corresponding to the quadratic function
\begin{equation}
    \sigma_T(k'[\mu_\circ^\vee]) = -\frac{1}{2}\moy{k'\mu^\vee_\circ{}^\ast,k'\mu_\circ^\vee} = -\frac{k'{}^2}{2}\tr\mu^\vee_\circ\mu^\vee_\circ,
\end{equation}
that is shifting $\theta$ by $2\pi$ is equivalent to adding the coupling $2\pi \cP_{\s_T}(b_\ttm)$. Identifying $\Gamma_\ttm$ with the integers modulo $k$ using the basis we chose earlier, we find $\sigma_T=(-\Delta p)\,\sigma_\circ$, where
\begin{equation}
    \Delta p \frac{1-k^2}{2k} = \frac{k'{}^2}{2}\tr\mu^\vee_\circ\mu^\vee_\circ \pmod{1}~,
\end{equation}
and from this we solve for $\Delta p$, yielding
\beq \Delta p = kk'{}^2\tr\mu^\vee_\circ\mu^\vee_\circ = \frac{M^2}{k}\tr\mu^\vee_\circ\mu^\vee_\circ ~, \label{eqn.deltap} \eeq
with the above equality taken modulo $2k$ for $k$ even and modulo $k$ for $k$ odd.\footnote{
To derive \eqref{eqn.deltap} for $k$ odd, we have made use of the fact that $kk'{}^2\tr\mu_0^\vee\mu_0^\vee$ is always an even integer.}

To summarize, there is a duality \eqref{T-duality}, which we call the $T$ transformation, between gauge theories labelled by the following continuous and discrete theta parameters
\begin{equation}
    (\theta+2\pi,p) \sim (\theta,p+(M^2/k)\tr \mu_\circ^\vee\mu_\circ^\vee)~. \label{identification}
\end{equation}
The discrete parameter $p$ is defined modulo $2k$ for even $k$, and modulo $k$ for odd $k$~\cite{Kapustin:2014gua}.

In section \ref{sec:non-abelian}, we will explain how $\Delta p = (M^2/k)\tr\m^\vee_\circ\m^\vee_\circ$ can be obtained by looking at the inverse Cartan matrix. We will also calculate $\Delta p$ for simple Lie groups, finding agreement with~\cite{Cordova:2019uob}. For the D$_N$ series with $N$ even, where $\Gamma_\ttm$ may not be cyclic, refer to section \ref{sec:D-series}.
\section{Non-Abelian Theories \label{sec:non-abelian}}
\subsection{\texorpdfstring{A Series: $\mathfrak{su}(N)$}{A Series: su(N)} \label{sec:su(n)}}
In this section we start the classification of line operators, with the Lie algebra $\mathfrak{su}(N)$ which has rank $N-1$ and the following Dynkin diagram
\[  \dynkin{A}{} \]
where each of the $N-1$ nodes correspond to the fundamental weights, denoted $\n_1, ..., \n_{N-1}$. The coweight lattice $\Lambda_\ttcw$ is generated by the fundamental coweights $\m^\vee_1, ...,\m^\vee_{N-1}$. (See Appendix \ref{app:lie.algebra} for our conventions with Lie algebras.) As mentioned in section \ref{sec:rules}, the weight of the fundamental 't Hooft line $T$ should be a generator of $Z(SU(N))=\Gamma_\ttm\cong \mathbb Z_N$, which we choose to be $\m^\vee_\circ := \m^\vee_1$. Having fixed this, a choice of the fundamental Wilson line satisfying \eqref{WT lines basis} is $\nu_\circ:=\nu_{N-1}$. Note that since
\begin{equation}
    \moy{\nu_i,\mu^\vee_j} = (C^{-1})_{ij}~,
\end{equation}
we can find appropriate choices of fundamental Wilson and 't Hooft lines by looking at the inverse of the Cartan matrix.

The dyonic Wilson-'t Hooft line $W^nT^m$ has weights $(n\n_{N-1},m\m^\vee_1)\in \Lambda_\ttw\times\Lambda_\ttcw$. Since $W^{2N}$ and $T^{2N}$ belong to twice the root $2\Lambda_\ttr$ and coroot $2\Lambda_\ttcr$ lattices, it is sufficient for our purposes to consider the mod $2N$ reduction of the integers $n$ and $m$. Therefore we label the line $W^nT^m$ by
\begin{equation}
    (n,m) \in \mathbb{Z}_{2N} \times \mathbb{Z}_{2N} \subset \tilde{\Lambda}~. \label{su(n) WT labels}
\end{equation}
A further mod $N$ quotient of $(n,m)$ yields the one-form symmetry charges
\begin{equation}
    (n\pmod{N},m\pmod{N}) \in \Lambda = \hat Z(SU(N))\times Z(SU(N))~.
\end{equation}

Let us find the Dirac quantization conditions \eqref{DSZ non-abelian} in terms of the integers \eqref{su(n) WT labels} by calculate the pairing between two dyonic lines of the form $W^n T^m$ and $W^{n'} T^{m'}$
\begin{equation}
    \big\langle (n,m), (n',m') \big\rangle_\mathrm{D} \equiv \frac{1}{N}\left(nm'-n'm\right) \pmod{2\mathbb Z}~. \label{su(n) pairing}
\end{equation}
Note that if we had defined $\n_1$ and $\m^\vee_1$ as the basis for $W$ and $T$, there would have been an extra factor of $N-1$, since $\langle \n_1 , \m^\vee_1 \rangle = (N-1)/N$. Requiring the RHS of \eqref{su(n) pairing} to be an integer, we get different solutions that are labeled by two integers $k$ and $q$ \cite{Aharony:2013hda}; the integer $k$ is a divisor of $N$, i.e. $N=kk'$ for some integer $k'$. The charge lattice of the theory $\mathcal{L}_{k,q}\subset \mathbb Z_{N} \times \mathbb Z_{N}$, is then generated by the lines
\begin{equation}
W^k \equiv (k,0) \quad \text{and} \quad W^qT^{k'} \equiv (q,k') ~, \label{su(n) spin sol}
\end{equation}
where $q=0,1,\dots, k-1$.\footnote{One can show that the charge lattice $\mathcal{L}_{k,q}$ is isomorphic to $\mathbb{Z}_{gcd(k,k',q)} \times \mathbb{Z}_{N/gcd(k,k',q)}$ \cite{Gaiotto:2014kfa}.} Now we can give a microscopic description of a theory with such set of lines. The value of $k$ fixes the gauge group to be $SU(N)/\mathbb{Z}_k$, and for spin theories $q$ can be identified with the theta parameter $p$ defined in \eqref{B cup B} as $q\equiv p \pmod{k}$. However for non-spin theories, we also need to determine the spin of the generating lines in \eqref{su(n) spin sol}.

We denote the theory where the generating electric line $W^k$ is bosonic and the magnetic generating line has weights and spin $(q,k')_s$, as $(SU(N)/{\mathbb{Z}_k})_{q,s}$. For theories where the generating electric line is fermionic, we denote the theory as $(\text{Spin-}SU(N)/{\mathbb{Z}_k})_{q,s}$ where again the other generating line is $(q,k')_s$. In summary we have the following possibilities with the following generating lines for non-spin theories:
\begin{align}
    \left( SU(N)/{\mathbb{Z}_k} \right)_{q,s}&: (k,0)_\ttb, \quad (q,k')_s~, \nonumber\\
    \left(\text{Spin-}SU(N)/{\mathbb{Z}_k}\right)_{q,s}&: (k,0)_\ttf, \quad (q,k')_s \label{su(n) non-spin sols} ~.
\end{align}
However, not all these choices are consistent. As stated in rule \ref{rule number 2} in section \ref{sec:rules}, lines labelled by just a root or a coroot must be bosonic. For instance, for $N$ odd, $k=1$ and $q=0$, the gauge group is $SU(N)$ and the fundamental Wilson line with weight $(\n_{N-1},0)$ cannot be fermionic; this is because fusing $N$ of them, according to \eqref{spin label}, results in a fermionic line labelled by a root, and hence is not allowed. We will find the consistency conditions below and demonstrate the details with a few examples. But before that, let us discuss the theta parameters in these theories.

Consider the $SU(N)/{\mathbb Z_k}$ theory whose set of lines and their spins are
\begin{equation}
    \leftidx{_{s'}}{{\left(SU(N)/{\mathbb Z_k}\right)}}{_{q,s}}: (k,0)_{s'}, \quad (q,k')_s~.
\end{equation}
Here the label $s'$ denotes $SU(N)/{\mathbb{Z}_k}$ for $s'=0$ and $(\text{Spin-}SU(N)/{\mathbb{Z}_k}$ for $s'=1$, see \eqref{su(n) non-spin sols}; these two notations are used interchangeably in this article. We claim that such theory can be obtained from ${\left(SU(N)/{\mathbb Z_k}\right)}_{0,\ttb}$ theory by adding the coupling \eqref{eqn.disctheta2} associated with the quadratic refinement 
\begin{equation}
    \sigma(nk+mq,mk') \equiv q\frac{m^2}{2k} + s \frac m2 + s'\frac n2 \equiv (q+sk)\frac{m^2}{2k} + s'\frac{n^2}{2} \pmod{\mathbb{Z}} ~,
\end{equation}
where $(nk+mq,mk') \in \mathbb{Z}_N \times \mathbb{Z}_N$ are the electric and magnetic charges. This simply follows from \eqref{eqn.tildesigma} by noting that the line $(nk+mp,mk')_{s''}$ can be obtained by fusing $(nk,0)_{ns'}$ with $(mq,mk')_{ms}$ and thus has spin $s'' \equiv ms+ns'+nm \pmod{2\mathbb Z}$. Such quadratic function induces a coupling of the form
\begin{equation}
    2\pi i\int_\mathcal{M} \left( \frac{q+sk}{2k}\mathcal{P}(b_\ttm ) + \frac{s'}{2} b_\tte \cup b_\tte \right)~, \label{generalized coupling SU(N)}    
\end{equation}
where the pair $(b_\tte,b_\ttm)$ essentially measure the integers $n$ and $m$. As explained in section \ref{anomalies.and.symmetries}, after adding such coupling the one-form symmetry extension changes which puts some constraints on the symmetry generators which restricts the coefficients in \eqref{generalized coupling SU(N)}. For instance we should check that $\sigma$ is a well-defined function; that is if we shift the electric and magnetic charges by $N$ it does not change mod $\mathbb Z$. One can easily see that this gives the relations
\begin{equation}
    ks + (k+s')q \equiv k's' \equiv 0 \pmod{2\mathbb Z}~. \label{consistency conditions for SU(N)}
\end{equation}
For the sake of completeness let us rederive these relations by the consistency conditions for the spin labels. Take $(k,0)_{s'}$ line and fuse it $k'$ times with it self to get $(N,0)_{k's'}$ which is a pure root and must be bosonic hence the first relation. Now fusing the line $(qk,0)_{qs'}$ with $(kq,kk')_{ks}$ gives $(0,N)_{qs'+ks+kq}$ which is bosonic hence the second relation. Therefore, we get the following consistent theories based on Lie group $\mathfrak{su}(N)$ on non-spin manifolds
\begin{align}
    \left( SU(N)/{\mathbb{Z}_k} \right)_{q,s}&: \quad \text{for} \quad k(q+s) \in {2\mathbb Z}~, \nonumber\\
    \left(\text{Spin-}SU(N)/{\mathbb{Z}_k}\right)_{q,s}&:  \quad \text{for} \quad  ks+(k+1)q,k' \in {2\mathbb Z}~.
\end{align}

Now we find the relation between the $T$-transformation and adding the coupling \eqref{generalized coupling SU(N)} for these theories. For the case where $s'=0$ and the basic Wilson line is bosonic, we have the usual theta parameter $p$ in \eqref{B cup B} and the identification
\begin{equation}
    q + ks \equiv (1-k^2)p \pmod{2k \mathbb Z}~, \label{q s and p}
\end{equation}
which automatically satisfies \eqref{consistency conditions for SU(N)}. A result of this relation is the fact that for even $k$, shifting $p$ by $k$ changes the spin of the basic 't Hooft line \cite{Hsin:2018vcg}, hence $p$ is $2k$ periodic. Furthermore since the basic 't Hooft line is $T^{k'}$, following the discussion around \eqref{identification}, we have $\Delta p = k({k'}\m^\vee_1,{k'}\m^\vee_1)=k'(N-1)$ and hence we get the following identification of the theta parameters
\begin{equation}
    (\theta+2\pi,p) \sim \left(\theta, p +k'(N-1) \right)~, \label{su(n) identification}
\end{equation}
where $p \in \mathbb Z_{k}$ for odd $k$ and $p \in \mathbb Z_{2k}$ for even $k$. Thus shifting $p$ by one and the $T$-transformation is equivalent to the following maps on the weights and spins
\begin{align}
    p \to p+1: (n,k'm)_{s_0} &\mapsto (n+(1-k^2)m,k'm)_{s_0}~, \label{p to p+1 su(n)}\\
    \theta \to \theta+2\pi: (n,k'm)_{s_0} &\mapsto (n+k'(N-1)m,k'm)_{s_0}~. \label{SU(N) T-transf}
\end{align}
Therefore we get the relation
\begin{equation}
    \left(SU(N)/{\mathbb{Z}_k}\right)_{q,s}^{\theta+2\pi} = \left(SU(N)/{\mathbb{Z}_k}\right)_{q+k'(N-1),s}^\theta~, \label{q,s vs theta}
\end{equation}
thus for $k$ even $\theta$ is $4\pi k/{{\text{gcd}(2k,k')}}$ periodic, and for odd $k$ it is $2\pi k/{{\text{gcd}(k,k')}}$ periodic.

For the $\text{Spin-}SU(N)/{\mathbb{Z}_k}$ theories \footnote{For the microscopic description of the $\text{Spin-}SU(N)/{\mathbb{Z}_k}:= \text{Spin}(4) \times_{\mathbb Z_2} SU(N)/{\mathbb{Z}_k}$ theory, we need a $\text{Spin}(4) \times_{\mathbb Z_2} SU(N)/{\mathbb{Z}_k}$ connection defined on a manifold which has $\text{Spin-}SU(N)/{\mathbb{Z}_k}$ structure. Where such structure is an extension of the $SO(N)$ frame bundle by the gauge $SU(N)/{\mathbb{Z}_k}$-bundle \cite{Wang:2018qoy}.} where $s'=1$, however the coupling \eqref{B cup B} does not make sense for arbitrary values of $p$ and we should really think of the coupling \eqref{generalized coupling SU(N)} instead, where from \eqref{consistency conditions for SU(N)} we have the relation $k' \equiv ks+(k+1)q \equiv 0 \pmod{2\mathbb Z}$. When $k$ is odd, $s$ must be even and we have the map
\begin{equation}
    q \to q+1: (n,k'm)_{s_0} \mapsto (n+m,k'm)_{s_0}~.
\end{equation}
However when $k$ is even, $q$ must be even too and we can only shift $q$ by an even number. In both cases by comparing the above map with the $T$-transformation we get the identification $(\theta+2\pi,q) \sim (\theta,q+k'(N-1))$. However, note that shifting $q$ by $k$ does not change the spins and therefore $q$ is only defined mod $k\mathbb Z$, hence for both $k$ even and odd $\theta$ is $2\pi k/{{\text{gcd}(k,k')}}$ periodic.
\subsection*{\texorpdfstring{$ \fg=\mathfrak{su}(2)$}{su(2)}}
Let us begin with the simplest example of the $\mathfrak{su}(2)$ algebra. This algebra has only one simple root and therefore
\begin{equation}
\n_\circ=\n_1 \quad \text{and} \quad \m^\vee_\circ=\m^\vee_1~.
\end{equation}
The Dirac quantization condition for the lines $W^nT^m$ and $W^{n'}T^{m'}$ is given by
\begin{equation}
    \langle (n,m), (n',m')\rangle_\mathrm{D} = \frac{1}{2}\left(nm'-n'm \right) \equiv 0 \pmod{\mathbb Z}~. \label{su(2) dsz}
\end{equation}
For spin theories only the charges $(n,m)$ mod $2$--modulo the root lattices--matters and we get three solutions by solving \eqref{su(2) dsz}. In each case the charge lattice $\mathcal{L} \subset \Lambda$ is isomorphic to $\mathbb Z_2$ and hence is generated by a single line operator as \cite{Aharony:2013hda}
\begin{equation}
SU(2):(1,0), \quad SO(3)_0: (0,1), \quad SO(3)_1:(1,1)~.
\end{equation}
The $T$-transformation \eqref{T transformation} acts as 
\begin{equation}
T:(n,m)_s\mapsto (n-m,m)_s~, \label{SU(2) Witten}
\end{equation}
and leaves the $SU(2)$ charge lattice invariant, while mapping $SO(3)_0$ to $SO(3)_1$.

\begin{figure}[ht]
     \centering
     \begin{subfigure}{0.4\textwidth}
         \centering  
         \begin{tikzpicture}[scale=1] 
 		\draw[-,black] (-1.5,0) -- (1.5,0);
 		\draw[-,black] (0,-2.5) -- (0,2.5);
 		\filldraw[black] (0,0) circle (2pt);
		\filldraw[black] (1,0) circle (2pt) node[anchor=north west] {b};
		\filldraw[black] (-1,0) circle (2pt) node[anchor=north west] {b};
		\filldraw[black,fill=white] (0,1) circle (2pt);
 		\filldraw[black,fill=white] (0,-1) circle (2pt);
 		\filldraw[black,fill=white] (1,1) circle (2pt);
 		\filldraw[black,fill=white] (1,-1) circle (2pt);
 		\filldraw[black,fill=white] (-1,1) circle (2pt);
 		\filldraw[black,fill=white] (-1,-1) circle (2pt);
 		\filldraw[black] (0,2) circle (2pt) node[anchor=north west] {b};
 		\filldraw[black] (0,-2) circle (2pt) node[anchor=north west] {b};
 		\filldraw[black] (1,2) circle (2pt) node[anchor=north west] {f};
 		\filldraw[black] (1,-2) circle (2pt) node[anchor=north west] {f};
 		\filldraw[black] (-1,2) circle (2pt) node[anchor=north west] {f};
 		\filldraw[black] (-1,-2) circle (2pt) node[anchor=north west] {f};
    	\end{tikzpicture}      
         \caption{$SU(2)$}
     \end{subfigure}
     \begin{subfigure}{0.4\textwidth}
         \centering         
          \begin{tikzpicture}[scale=1] 
 		\draw[-,black] (-1.5,0) -- (1.5,0);
 		\draw[-,black] (0,-2.5) -- (0,2.5);
 		\filldraw[black] (0,0) circle (2pt);
		\filldraw[black] (1,0) circle (2pt) node[anchor=north west] {f};
		\filldraw[black] (-1,0) circle (2pt) node[anchor=north west] {f};
		\filldraw[black,fill=white] (0,1) circle (2pt);
 		\filldraw[black,fill=white] (0,-1) circle (2pt);
 		\filldraw[black,fill=white] (1,1) circle (2pt);
 		\filldraw[black,fill=white] (1,-1) circle (2pt);
 		\filldraw[black,fill=white] (-1,1) circle (2pt);
 		\filldraw[black,fill=white] (-1,-1) circle (2pt);
 		\filldraw[black] (0,2) circle (2pt) node[anchor=north west] {b};
 		\filldraw[black] (0,-2) circle (2pt) node[anchor=north west] {b};
 		\filldraw[black] (1,2) circle (2pt) node[anchor=north west] {b};
 		\filldraw[black] (1,-2) circle (2pt) node[anchor=north west] {b};
 		\filldraw[black] (-1,2) circle (2pt) node[anchor=north west] {b};
 		\filldraw[black] (-1,-2) circle (2pt) node[anchor=north west] {b};
    	\end{tikzpicture}      
         \caption{$\text{Spin-}SU(2)$}
     \end{subfigure}
 \caption{$SU(2)$ lattices}
 \label{fig:SU(2)}
\end{figure}
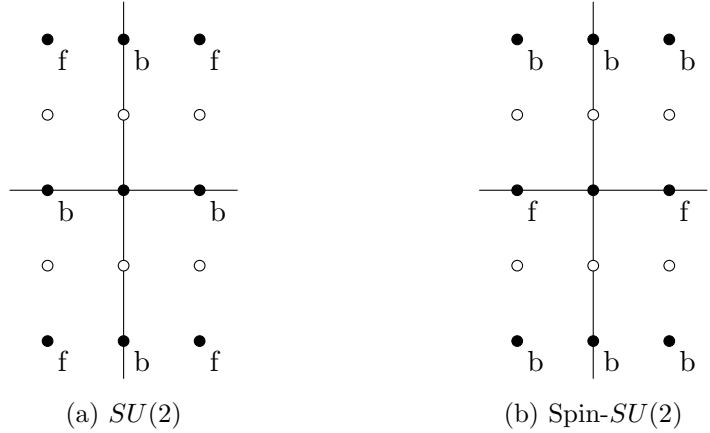

Going to the case of oriented theories without spin-structure, we label the lines by their weights in the quotient \eqref{quotient} which can be identified with mod $4$ reduction of $n$ and $m$, i.e.  $(n,m) \in \Lambda_\ttw/2\Lambda_\ttr \times \Lambda_\ttcw/2\Lambda_\ttcr  \cong \mathbb Z_4 \times \mathbb Z_4$. For instance, the lines $T$ and $W^2T$ have the same one-form charge in $\Lambda$, but they are distinguishable since they have different spins. For $SU(2)$ we have two lattices, one where the generating line is $(1,0)_\ttb$--a boson--and one where it is a fermion, i.e. $(1,0)_\ttf$. As we mentioned in section \ref{sec:rules} we require that the adjoint line $(0,2)$ is always bosonic; this along with the fusion rule \eqref{spin label} fixes the spin of all the lines completely. For each case, we get the following set of weights and spins
\begin{align}
    SU(2) &: (\pm1,0)_\ttb, (0,2)_\ttb, (\pm1,2)_\ttf, (2,0)_\ttb~, (2,2)_\ttb, (0,0)_\ttb~,\\
    \text{Spin-}SU(2) &: (\pm1,0)_\ttf, (0,2)_\ttb, (\pm1,2)_\ttb, (2,0)_\ttb~, (2,2)_\ttb, (0,0)_\ttb~;
\end{align}
which are also depicted in the figure \ref{fig:SU(2)}. The $T$-transformation leaves both lattices invariant and therefore $\theta$ is $2\pi$ periodic for both.

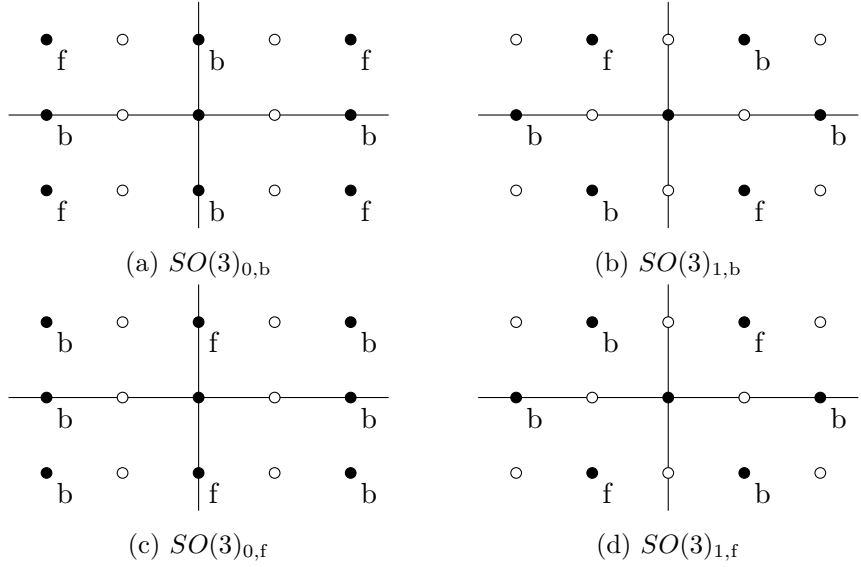
\begin{figure}[ht]
     \centering
     \begin{subfigure}{0.4\textwidth}
         \centering  
         \begin{tikzpicture}[scale=1] 
 		\draw[-,black] (-2.5,0) -- (2.5,0);
 		\draw[-,black] (0,-1.5) -- (0,1.5);
 		\filldraw[black] (0,0) circle (2pt);
		\filldraw[black] (0,1) circle (2pt) node[anchor=north west] {b};
		\filldraw[black] (0,-1) circle (2pt) node[anchor=north west] {b};
		\filldraw[black,fill=white] (1,0) circle (2pt);
 		\filldraw[black,fill=white] (-1,0) circle (2pt);
 		\filldraw[black,fill=white] (1,1) circle (2pt);
 		\filldraw[black,fill=white] (-1,1) circle (2pt);
 		\filldraw[black,fill=white] (1,-1) circle (2pt);
 		\filldraw[black,fill=white] (-1,-1) circle (2pt);
 		\filldraw[black] (2,0) circle (2pt) node[anchor=north west] {b};
 		\filldraw[black] (-2,0) circle (2pt) node[anchor=north west] {b};
 		\filldraw[black] (2,1) circle (2pt) node[anchor=north west] {f};
 		\filldraw[black] (-2,1) circle (2pt) node[anchor=north west] {f};
 		\filldraw[black] (2,-1) circle (2pt) node[anchor=north west] {f};
 		\filldraw[black] (-2,-1) circle (2pt) node[anchor=north west] {f};
    	\end{tikzpicture}      
        \subcaption{$SO(3)_{0,\ttb}$}
     \end{subfigure}
     \begin{subfigure}{0.4\textwidth}
         \centering  
         \begin{tikzpicture}[scale=1] 
 		\draw[-,black] (-2.5,0) -- (2.5,0);
 		\draw[-,black] (0,-1.5) -- (0,1.5);
 		\filldraw[black] (0,0) circle (2pt);
		\filldraw[black] (1,1) circle (2pt) node[anchor=north west] {b};
		\filldraw[black] (-1,-1) circle (2pt) node[anchor=north west] {b};
		\filldraw[black,fill=white] (1,0) circle (2pt);
 		\filldraw[black,fill=white] (-1,0) circle (2pt);
 		\filldraw[black,fill=white] (2,1) circle (2pt);
 		\filldraw[black,fill=white] (0,1) circle (2pt);
 		\filldraw[black,fill=white] (0,-1) circle (2pt);
 		\filldraw[black,fill=white] (-2,-1) circle (2pt);
 		\filldraw[black,fill=white] (2,-1) circle (2pt);
 		\filldraw[black,fill=white] (-2,1) circle (2pt);
 		\filldraw[black] (2,0) circle (2pt) node[anchor=north west] {b};
 		\filldraw[black] (-2,0) circle (2pt) node[anchor=north west] {b};
 		\filldraw[black] (-1,1) circle (2pt) node[anchor=north west] {f};
 		\filldraw[black] (1,-1) circle (2pt) node[anchor=north west] {f};
    	\end{tikzpicture}      
        \subcaption{$SO(3)_{1,\ttb}$}
     \end{subfigure}

     \begin{subfigure}{0.4\textwidth}
         \centering         
          \begin{tikzpicture}[scale=1] 
 		\draw[-,black] (-2.5,0) -- (2.5,0);
 		\draw[-,black] (0,-1.5) -- (0,1.5);
 		\filldraw[black] (0,0) circle (2pt);
		\filldraw[black] (0,1) circle (2pt) node[anchor=north west] {f};
		\filldraw[black] (0,-1) circle (2pt) node[anchor=north west] {f};
		\filldraw[black,fill=white] (1,0) circle (2pt);
 		\filldraw[black,fill=white] (-1,0) circle (2pt);
 		\filldraw[black,fill=white] (1,1) circle (2pt);
 		\filldraw[black,fill=white] (-1,1) circle (2pt);
 		\filldraw[black,fill=white] (1,-1) circle (2pt);
 		\filldraw[black,fill=white] (-1,-1) circle (2pt);
 		\filldraw[black] (2,0) circle (2pt) node[anchor=north west] {b};
 		\filldraw[black] (-2,0) circle (2pt) node[anchor=north west] {b};
 		\filldraw[black] (2,1) circle (2pt) node[anchor=north west] {b};
 		\filldraw[black] (-2,1) circle (2pt) node[anchor=north west] {b};
 		\filldraw[black] (2,-1) circle (2pt) node[anchor=north west] {b};
 		\filldraw[black] (-2,-1) circle (2pt) node[anchor=north west] {b};
    	\end{tikzpicture}        
        \subcaption{$SO(3)_{0,\ttf}$}
     \end{subfigure}
     \begin{subfigure}{0.4\textwidth}
         \centering  
         \begin{tikzpicture}[scale=1] 
 		\draw[-,black] (-2.5,0) -- (2.5,0);
 		\draw[-,black] (0,-1.5) -- (0,1.5);
 		\filldraw[black] (0,0) circle (2pt);
		\filldraw[black] (1,1) circle (2pt) node[anchor=north west] {f};
		\filldraw[black] (-1,-1) circle (2pt) node[anchor=north west] {f};
		\filldraw[black,fill=white] (1,0) circle (2pt);
 		\filldraw[black,fill=white] (-1,0) circle (2pt);
 		\filldraw[black,fill=white] (2,1) circle (2pt);
 		\filldraw[black,fill=white] (0,1) circle (2pt);
 		\filldraw[black,fill=white] (0,-1) circle (2pt);
 		\filldraw[black,fill=white] (-2,-1) circle (2pt);
 		\filldraw[black,fill=white] (2,-1) circle (2pt);
 		\filldraw[black,fill=white] (-2,1) circle (2pt);
 		\filldraw[black] (2,0) circle (2pt) node[anchor=north west] {b};
 		\filldraw[black] (-2,0) circle (2pt) node[anchor=north west] {b};
 		\filldraw[black] (-1,1) circle (2pt) node[anchor=north west] {b};
 		\filldraw[black] (1,-1) circle (2pt) node[anchor=north west] {b};
    	\end{tikzpicture}      
        \subcaption{$SO(3)_{1,\ttf}$}
     \end{subfigure}
 \caption{$SO(3)$ lattices}
 \label{fig:SO(3)}
\end{figure}

Similarly for each $SO(3)$ we get two lattices, one where the generating line is bosonic and fermionic. In total we get four lattices shown in figure \ref{fig:SO(3)}.
All of these lattices are related by the $T$-transformation as
\begin{equation}
    SO(3)_{q,s}^{\theta+2\pi} = SO(3)_{q+1,s}^{\theta}~,
\end{equation}
hence $\theta$ is $8\pi$ periodic. From \eqref{q s and p} we have $p \equiv q + 2s \pmod{\mathbb Z}$, thus we get the identification $(\theta+2\pi,p) \sim (\theta, p +1)$. We also have the relation
\begin{equation}
    SO(3)_{q,s}^\theta = SO(3)_{q+2,s+1}^{\theta}~,
\end{equation}
which can be verified by either looking at the weight lattices of these theories and match their line operators, looking at the relation $p \equiv q + 2s \pmod{\mathbb Z}$, or looking at the coupling \eqref{B cup B} for $p=2$. Using the Wu formula $\frac12 b_\ttm \cup b_\ttm  \equiv \frac12 b_\ttm \cup w_2(T\mathcal{M}) \pmod{\mathbb Z}$, this coupling is equivalent to adding $\pi i \int_{\mathcal{M}} b_\ttm \cup w_2(T\mathcal{M})$, which attaches a $w_2(T\mathcal{M})$ surface to the fundamental 't Hooft line and hence changes its spin \cite{Thorngren:2014pza,Cordova:2018acb,Wang:2018qoy,Hsin:2018vcg}. To see these relations explicitly, let us take a closer look at how the $T$-transformation \eqref{SU(2) Witten} maps the line operators of these four theories to each other:
\begin{equation}
\begin{array}{ccccccc}
SO(3)_{0,\ttb}  &\xmapsto{T}& SO(3)_{1,\ttf} &\xmapsto{T}& SO(3)_{0,\ttf} &\xmapsto{T}& SO(3)_{1,\ttb} \\ \hline
(0,\pm1)_\ttb         &\mapsto& ~(\mp1,\pm1)_\ttb    &\mapsto& ~(2,\pm1)_\ttb       &\mapsto& ~(\pm1,\pm1)_\ttb\\
(2,\pm1)_\ttf         &\mapsto& (\pm1,\pm1)_\ttf      &\mapsto& (0,\pm1)_\ttf       &\mapsto& (\mp1,\pm1)_\ttf\\
(2,0)_\ttb            &\mapsto& (2,0)_\ttb            &\mapsto& (2,0)_\ttb          &\mapsto& (2,0)_\ttb\\
(0,2)_\ttb            &\mapsto& (2,2)_\ttb            &\mapsto& (0,2)_\ttb          &\mapsto& (2,2)_\ttb\\
(2,2)_\ttb            &\mapsto& (0,2)_\ttb            &\mapsto& (2,2)_\ttb          &\mapsto& (0,2)_\ttb \label{lines of SO(3)}
\end{array}  
\end{equation}
\subsection*{\texorpdfstring{$\fg=\mathfrak{su}(4)$}{su(4)}}
As the last example of $\mathfrak{su}(N)$, in this section we study lines operators of $\mathfrak{su}(4)$ theories. The weights of the basic Wilson and 't Hooft lines are given by $\n_\circ := \n_3$ and $\m^\vee_\circ:=\m^\vee_1$. Here we only consider dyonic lines of the form $W^nT^m$, denoted by their weights $(n,m)\in \mathbb{Z}_8 \times \mathbb{Z}_8 \subset \tilde{\Lambda}$ in the quotient \eqref{quotient}. From \eqref{su(n) spin sol}, on spin manifolds different charge lattices are generated by the following lines
\begin{equation}
SU(4): \, (1,0),\,(0,4), \quad (SU(4)/\mathbb{Z}_2)_{q=0,1}: \, (2,0), \,(q,2), \quad PSU(4)_{q=0,1,2,3}:\, (4,0),\,(q,1)~. \nonumber
\end{equation}
Let us first review the results of \cite{Aharony:2013hda} for these theories on spin manifolds. For the case of $SU(4)$, there is only one theory and $\theta$ is $2\pi$ periodic. For gauge group $SO(6)=SU(4)/\mathbb{Z}_2$, the two solutions with $p=0$ and $p=1$ are not related by the $T$-transformation and $\theta$ is $2\pi$ periodic. Finally for $PSU(4)$, all the four theories are related by the $T$-transformation and hence $\theta$ is $8\pi$ periodic.

Moving to non-spin manifolds, we need to also determine the spin of the generating lines of the above charge lattices. For $SU(4)$ there are two theories one where the basic Wilson line $W(\gamma)$ is bosonic and one where it is fermionic. Both are consistent since the adjoint lines $W^4$ and $T^4$ are bosonic. Thus we get two solutions denoted as $SU(4)$ and $\text{Spin-}SU(4)$. From \eqref{SU(N) T-transf}, we see that for both lattices $\theta$ is $2\pi$ periodic.

For $SO(6)_0$ both the generating $(2,0)$ and $(0,2)$ lines can be either bosonic or fermionic and we get four theories. Whereas for $SO(6)_1$, the line $(2,0)$ can only be bosonic; this is because it can be obtained by fusing the line $(1,4)$ with itself and therefore must be bosonic. Hence there is no $\text{Spin-}SO(6)_{1,s}$ solution and in total we get six theories. From \eqref{q s and p}, we have $q + 2s \equiv p \pmod{4}$. To verify this relation, let us study the action of shifting $p$ by one--given in \eqref{p to p+1 su(n)}--on the line operators of different $SO(6)$ theories
\begin{equation}
    \begin{array}{ccccccc}
    SO(6)_{0,\ttb} &\mapsto& SO(6)_{1,\ttb} &\mapsto& SO(6)_{0,\ttf} &\mapsto& SO(6)_{1,\ttf} \\ \hline
    (0,2)_\ttb     &\mapsto& (1,2)_\ttb     &\mapsto& (2,2)_\ttb     &\mapsto& (3,2)_\ttb \\
    (2,0)_\ttb     &\mapsto& (2,0)_\ttb     &\mapsto& (2,0)_\ttb     &\mapsto& (2,0)_\ttb \\
    (2,2)_\ttf     &\mapsto& (3,2)_\ttf     &\mapsto& (4,2)_\ttf     &\mapsto& (5,2)_\ttf \\
    (6,2)_\ttf     &\mapsto& (7,2)_\ttf     &\mapsto& (0,2)_\ttf     &\mapsto& (1,2)_\ttf \\
    (2,6)_\ttf     &\mapsto& (5,6)_\ttf     &\mapsto& (0,6)_\ttf     &\mapsto& (3,6)_\ttf 
    \end{array}
\end{equation}

There is also the identification the theta parameters as $(\theta+2\pi,p)\sim(\theta,p+2)$. Thus we get the action of the T-transformation on these theories as shown in \eqref{SO(6) T} and \eqref{spin SO(6) T}.
\begin{equation}
\begin{tikzcd}
	SO(6)_{0,\ttb} \arrow[r, leftrightarrow,"T"] & SO(6)_{0,\ttf}
\end{tikzcd}
\quad
\begin{tikzcd}
	SO(6)_{1,\ttb} \arrow[r, leftrightarrow,"T"] & SO(6)_{1,\ttf}
\end{tikzcd}
\label{SO(6) T}
\end{equation}
\begin{equation}
\begin{tikzcd}
	\text{Spin-}SO(6)_{\ttb} \arrow[loop left, leftrightarrow,"T"]
\end{tikzcd}
\quad
\begin{tikzcd}
	\text{Spin-}SO(6)_{\ttf} \arrow[loop right, leftrightarrow,"T"]
\end{tikzcd}
\label{spin SO(6) T}
\end{equation}
Hence $\theta$ is $4\pi$ periodic for the $SO(6)$ theories, and $2\pi$ for the Spin-$SO(6)$ theories.

For the $PSU(4)_q$ theories, the charge lattice is generated by $(q,1)$ which can have any spin. Thus we get eight different theories denoted as $PSU(4)_{q,s}$, where the generating line has weights and spin $(q,1)_s$. All of these eight theories are related by adding the coupling \eqref{B cup B}, where $p \equiv q + 4s \pmod{8}$. There is also the identification $(\theta+2\pi,p)\sim(\theta,p+3)$, thus all of these theories are also scanned by the $T$-transformation and we have the relations
\begin{equation}
    PSU(4)^{\theta+2\pi}_{q,s} = PSU(4)^{\theta}_{q+3,s},\quad PSU(4)^{\theta}_{q,\ttb} = PSU(4)^{\theta}_{q+4,\ttf}~.
\end{equation}

These theories have interesting 't Hooft anomalies which are discussed in section \ref{sec:background fields}. In particular, the $\text{Spin-}SO(6)_{\ttf}$ theory has the same gravitational anomaly as the all-fermion electrodynamics theory ${\text{Spin}^\mathbb C}_\ttf$.

\subsection{\texorpdfstring{C Series: $\mathfrak{sp}(N)$}{C Series: sp(N)}}
The universal covering group of $\fg=\mathfrak{sp}(N)$ is $Sp(N)$, whose center is isomorphic to $\bZ_2$. The Dynkin diagram of the $\mathfrak{sp}(N)$ algebra is
\[  \dynkin{C}{}\]
where the nodes correspond to the $N$ fundamental weights denoted as $\n_1, ..., \n_{N}$ with $\a_{N}$ the long root. The Langlands dual Lie algebra is $^L\fg=\mathfrak{so}(2N+1)$ with the Dynkin diagram
\[  \dynkin{B}{}\]
where the nodes correspond to the $N$ fundamental coweights $\m^\vee_1, ...,\m^\vee_{N}$ with $\a^\vee_N$ the short coroot. $[\m^\vee_{N}]$ generates the center $Z(Sp(N))$, so we choose $\m^\vee_\circ:=\m^\vee_{N}$ as the weight of the fundamental 't Hooft line $T$ and the fundamental representation $\n_\circ:=\n_1$ as the weight of the fundamental Wilson line $W$, since $\langle \n_1, \m^\vee_N \rangle = \frac12$. Labelling the dyonic lines $W^nT^m$ by $(n,m)$, the \eqref{DSZ non-abelian} condition becomes
\begin{equation}
    \big\langle (n,m),(n',m') \big\rangle_\mathrm{D} = \frac12 (nm'-n'm) = 0 \pmod{\mathbb Z}~.
\end{equation}
Following the same method as $\mathfrak{su}(N)$, we get the following theories, written below together with lines generating the set of allowed line operators:
\begin{align}
    Sp(N)&: (1,0)_\ttb, \quad (0,2)_\ttb~, \nonumber\\
    \text{Spin-}Sp(N)&: (1,0)_\ttf, \quad (0,2)_\ttb~,\\
    \left( Sp(N)/{\mathbb{Z}_2} \right)_{q,s}&: (2,0)_\ttb, \quad (q,1)_s~, \nn
\end{align}
where $q=0,1$. For the universal covering group $Sp(N)$, there is no possible discrete theta term and $\theta$ is $2\pi$ periodic. For the $Sp(N)/{\mathbb{Z}_2}$ theories, the periodicity of $\theta$ is increased, and according to \eqref{identification}, the $T$-transformation identifies theories differing by $\Delta p = 2 (\m^\vee_\circ,\m^\vee_\circ)= 2(C^{-1})_{NN}=N$ and the following continuous and discrete theta parameters are identified
\begin{equation}
    (\theta+2\pi,p)\sim(\theta,p+N)~,
\end{equation}
where $p \equiv q+2s \pmod{4}$, in agreement with \cite{Cordova:2019uob}. Thus, the various $Sp(N)/\bZ_2$ theories have the following relations
\begin{equation}
    \left( Sp(N)/{\mathbb{Z}_2} \right)_{q,s}^{\theta+2\pi} = \left( Sp(N)/{\mathbb{Z}_2} \right)_{q+N,s}^{\theta}, \quad \left( Sp(N)/{\mathbb{Z}_2} \right)_{q+2,\ttb}^{\theta} = \left( Sp(N)/{\mathbb{Z}_2} \right)_{q,\ttf}^{\theta}~.
\end{equation}
\subsection{\texorpdfstring{B Series: $\mathfrak{so}(2N+1)$, $N\geq 2$}{B Series: so(2N+1)}}
The classification of the lines for theories with $\mathfrak{so}(2N+1)$ algebra proceeds similarly to that of the Langlands dual algebra $\mathfrak{sp}(N)$, but with roots and coroots exchanged. The universal covering group is $Spin(2N+1)$, with center isomorphic to $\mathbb Z_2$. The Dynkin diagram is
\[  \dynkin{B}{}~. \]
We denote the simple roots by $\alpha_1,\ldots,\alpha_N$ with $\alpha_N$ the short root, and the corresponding fundamental weights by $\n_1, ..., \n_{N}$ and fundamental coweights by $\m^\vee_1, ...,\m^\vee_{N}$. Note that $[\m^\vee_1]$ generates the center $\Lambda_\ttcw/\Lambda_\ttcr = Z(Spin(2N+1))= \mathbb Z_2$, and $\nu_N$ has $1/2$ pairing with $\mu^\vee_1$, so we choose $\m^\vee_\circ:=\m^\vee_1$ and $\n_\circ:=\n_N$ as the weights of the fundamental 't Hooft and Wilson lines $T$ and $W$. Solving the quantization condition \eqref{DSZ non-abelian} yields the following theories
\begin{align}
    Spin(2N+1)&: (1,0)_\ttb, \quad (0,2)_\ttb~, \nonumber\\
    \Spin\text{-}Spin(2N+1)&: (1,0)_\ttf, \quad (0,2)_\ttb~,\\
    SO(2N+1)_{q,s}&: (2,0)_\ttb, \quad (q,1)_s~, \nn
\end{align}
for $q=0,1$. According to \eqref{identification}, the $T$-transformation identifies theories with the theta parameters
\begin{equation}
    (\theta+2\pi,p)\sim(\theta,p+2(C^{-1})_{11}) = (\theta,p+2)~,
\end{equation}
where, once again, $p \equiv q+2s \pmod{4}$. Thus, the various $SO(2N+1)$ theories have the following relations
\begin{equation}
    SO(2N+1)_{q,s}^{\theta+2\pi} = SO(2N+1)_{q+2,s}^{\theta} = SO(2N+1)_{q,s+1}^{\theta}~.
\end{equation}
\subsection{\texorpdfstring{D Series: $\mathfrak{so}(2N)$, $N\geq 3$}{D Series: so(2N)} }
\label{sec:D-series}
The universal covering group is $Spin(2N)$ and the Dynkin diagram is
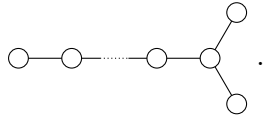
\[  \dynkin{D}{}~. \]
We denote the fundamental weights by $\n_1, ..., \n_{N-2}, \n_\tts, \n_\ttc$, where $\n_\tts$ and $\n_\ttc$ are the highest weights of the spinor and conjugate spinor representations respectively, and the corresponding fundamental coweights by $\m^\vee_1, ..., \m^\vee_{N-2}, \m^\vee_\tts, \m^\vee_\ttc$. The inverse of the Cartan matrix is 
\begin{align}
    C^{-1}_{ij} = \min\{i,j\}~, \quad\quad &{C^{-1}}_{i\tts} = {C^{-1}}_{i\ttc} = \frac i 2~, \nn \\
    {C^{-1}}_{\tts \tts}={C^{-1}}_{\ttc \ttc}=\frac{N}{4}~, \quad\quad &{C^{-1}}_{\tts \ttc}={C^{-1}}_{\ttc \tts}=\frac{N-2}{4}~, \label{inverse-cartan-matrix-of-Dn}
\end{align}
where $i,j=1,...,N-2$.

The classification of the lines for theories with $D$-series gauge groups differs qualitatively depending on the parity of $N$. For odd $N$, the center $Z(Spin(2N))=\Lambda_\ttcw/\Lambda_\ttcr$ is isomorphic to $\bZ_4$, and is generated by either $[\m^\vee_\tts]$ or $[\m^\vee_\ttc]$. For even $N$, the center is isomorphic to $\mathbb Z_2^\tts\times \mathbb Z_2^\ttc$, with generators $[\m^\vee_\tts]$ and $[\m^\vee_\ttc]$. We shall discuss each case separately.

\subsection*{Odd $N$}
We choose the coweight $\mu^\vee_\circ$ of the fundamental 't Hooft line $T$ to be $\m^\vee_\circ := \m^\vee_\tts$, and the weight $\nu_\circ$ of the fundamental Wilson line $W$ such that $\langle \n_\circ, \m^\vee_\circ \rangle = 1/4$.\footnote{
Unlike in the previous cases, there are no fundamental weights that satisfy this relation.
}
The condition \eqref{DSZ non-abelian} for the dyonic lines $W^nT^m$ and $W^{n'}T^{m'}$ then becomes
\begin{equation}
    \big\langle (n,m),(n',m') \big\rangle = \frac14 (nm'-n'm) = 0 \pmod{\mathbb Z}~.
\end{equation}
Labeling the lines with spins consistent with \eqref{spin label}, and solving \eqref{DSZ non-abelian}, we get the following theories
\begin{alignat}{2}
    Spin(2N)&: (1,0)_\ttb\,, \;(0,4)_\ttb\,, \quad \quad \quad \Spin\text{-}Spin(2N)&: (1,0)_\ttf\,, \;(0,4)_\ttb~, \nn \\
    SO(2N)_{q,s}&: (2,0)_\ttb\,, \;(q,2)_s\,, \quad \quad \quad \Spin\text{-}SO(2N)_{0,s}&: (2,0)_\ttf\,, \;(0,2)_s~, \\
    PSO(2N)_{q,s}&: (4,0)_\ttb\,, \;(q,1)_s~. \nn
\end{alignat}
For theories with gauge group $Spin(2N)$ and $\Spin$-$Spin(2N)$, the $\theta$ parameter is $2\pi$-periodic. For the $\left(\Spin(2N)/{\mathbb Z_k}\right)_{q,s}$ theories (with $k=2,4$), theta periodicity is enhanced. Following the discussion around \eqref{identification}, a $2\pi$ shift in the theta angle is dual to a discrete theta angle of
\beq \Delta p = kk'{}^2\tr\m^\vee_\circ\m^\vee_\circ = \frac{16}{k} C^{-1}_{\tts \tts}=\frac{4N}{k} \pmod{2k}. \eeq
Hence, the following theta parameters are identified
\begin{equation}
    (\theta+2\pi,p) \sim (\theta,p+4N/k)~,
\end{equation}
where $p \equiv q+sk \pmod{2k}$. $\theta$ is $4\pi$-periodic for the $SO(N)$ theories and $16\pi$-periodic for the $PSO(N)$ theories. The following theories are dual:
\begin{equation}
    SO(N)_{q,s}^{\theta+2\pi} = SO(N)_{q+2N,s}^{\theta}~, \qquad PSO(N)_{q,s}^{\theta+2\pi} = PSO(N)_{q+N,s}^{\theta}~.
\end{equation}
\subsection*{Even $N$}
When $N$ is even, the center of the universal covering group $Spin(2N)$ is $\bZ_2^\tts \times\bZ_2^\ttc$, which is novel, as all the $\pi_1(G)$ we have encountered thus far have been cyclic. The center is generated by the fundamental coweights corresponding to the spinor and conjugate spinor representations
\begin{equation}
    \m^{\vee}_{\circ1} := \m^\vee_\tts, \qquad \m^{\vee}_{\circ2} := \m^\vee_\ttc~,
\end{equation}
and we denote the 't Hooft lines with these coweights as $T_1$ and $T_2$ respectively. The weights of the fundamental Wilson lines $W_1$ and $W_2$ are chosen such that
\begin{equation}
    \langle \n_{\circ i}, \m_{\circ j}^\vee \rangle = \frac12\d_{ij}~.
\end{equation}
Denote the Wilson-'t Hooft line $W_1^{n_1}W_2^{n_2}T_1^{m_1}T_2^{m_2}$ by $(n_1,n_2;m_1,m_2)$. The quantization condition \eqref{DSZ non-abelian} reads
\begin{equation*}
    \big \langle (n_1,n_2;m_1,m_2), (n_1,n_2;m_1,m_2) \big \rangle = \frac{1}{2} (n_1 m_1' + n_2m_2' - n_1'm_1 - n_2'm_2) = 0 \pmod{\bZ}~. \label{d series dsz}
\end{equation*}
The following are the possible solutions of this condition. Next to each theory, we have given two lines generating the full set of line operators.
\begin{equation}
\begin{array}{rll}
    \leftidx{_{s_i'}}{Spin(2N)}{}&: (1,0;0,0)_{s_1'}\,, &\quad (0,1;0,0)_{s_2'}~, \\ 
    \leftidx{_{s'}}{SO(2N)}{_{q,s}}&: (1,-1;0,0)_{s'}\,, &\quad (q,0;1,1)_{s}~, \\ 
    \leftidx{_{s'}}{Ss(2N)}{_{q,s}}&: (0,1;0,0)_{s'}\,, &\quad (q,0;1,0)_{s}~, \\ 
    \leftidx{_{s'}}{Sc(2N)}{_{q,s}}&: (1,0;0,0)_{s'}\,, &\quad (0,q;0,1)_{s}~, \\ 
    PSO(2N)_{q_{ij},s_i}&: (q_{11},q_{21};1,0)_{s_1}\,, &\quad (q_{12},q_{22};0,1)_{s_2}~,
\end{array}  
\end{equation}
where each $q_{ij}=0$ or $1$, and $q_{12}=q_{21}$. We have defined the various gauge groups with algebra $\mathfrak{so}(2N)$ following the notation of \cite{Aharony:2013hda}, by taking quotients of $Spin(2N)$ by the subgroups of its center $\Lambda_\ttcw/\Lambda_\ttcr=\bZ_2^\tts \times \bZ_2^\ttc$. To wit, $SO(2N):=Spin(2N)/\bZ_2^\mathrm{v}$, $Ss(2N):=Spin(2N)/\bZ_2^
\tts$, $Sc(2N):=Spin(2N)/\bZ_2^\ttc$ and $PSO(2N):=Spin(2N)/(\bZ_2^\tts\times\bZ_2^\ttc)$, where the subgroups $\bZ_2^\mathrm{v}$, $\bZ_2^\tts$ and $\bZ_2^\ttc$ are generated by the vector, spinor and conjugate spinor representations of the dual Lie algebra $^L\mathfrak{so}(2N)\approx\mathfrak{so}(2N)$ respectively. As before, the left subscript $s'$ denotes the spin of the fundamental Wilson line; for instance $\leftidx{_{1}}{SO(2N)}{}:=\Spin\text{-}SO(2N)$. Note that, as there are three $\bZ_2$ subgroups of $Z(Spin(2N))=\bZ_2^\tts \times \bZ_2^\ttc$, the notation $\Spin\text{-}Spin(2N)$ is ambiguous and it is more precise to use the notation with the left subscript.

We now investigate the relation between theta terms. The $SO(2N)$, $Ss(2N)$ and $Sc(2N)$ theories have $\Gamma_\ttm \cong \bZ_2$, and there is a single discrete theta term $p\equiv q+2s \pmod{4}$. For $SO(2N)$ theories, \eqref{identification} yields
\beq \Delta p = 2\tr(\m^\vee_\tts+\m^\vee_\ttc)(\m^\vee_\tts+\m^\vee_\ttc)=2 \pmod{4} \eeq
leading to the identification
\beq (\theta+2\pi,p) \sim (\theta,p+2) \eeq
and therefore
\begin{equation}
    \leftidx{_{s'}}{SO(2N)}{^{\theta+2\pi}_{q,s}} = \leftidx{_{s'}}{SO(2N)}{^{\theta}_{q+2,s}} = \leftidx{_{s'}}{SO(2N)}{^{\theta}_{q,s+1}}~.
\end{equation}
For the $Ss(2N)$ and $Sc(2N)$ theories,
\beq \Delta p = 2\tr\m^\vee_\tts\m^\vee_\tts =\frac{N}{2}\pmod{4} ~, \eeq
leading to the identification
\beq (\theta+2\pi,p) \sim (\theta,p+N/2) \eeq
and therefore
\begin{equation}
    \leftidx{_{s'}}{Ss(2N)}{^{\theta+2\pi}_{q,s}} = \leftidx{_{s'}}{Ss(2N)}{^{\theta}_{q+N/2,s}}, \quad \leftidx{_{s'}}{Sc(2N)}{^{\theta+2\pi}_{q,s}} = \leftidx{_{s'}}{Sc(2N)}{^{\theta}_{q+N/2,s}}~.
\end{equation}
Note that for all of these theories -- $SO(2N)$, $Ss(2N)$ and $Sc(2N)$ -- there is no relation between $s$, $q$ and $s'$ since the electric sector is decoupled from the magnetic sector, i.e. the basic Wilson line has zero pairing with the basic 't Hooft line. Furthermore the one-form symmetry group is the direct product of the $\mathbb Z_2$ electric and $\bZ_2$ magnetic symmetry groups.

For $PSO(2N)$ theories, the magnetic one-form symmetry group is $\widehat{\Gamma}_\ttm=\widehat{\bZ}_2^\tts\times\widehat{\bZ}_2^\ttc$. Here, we arrive at a novel situation not discussed in section \ref{sec:Line Operators}: $\Gamma_\ttm$ is not cyclic. In this case, on top of the usual $\bZ_4$-valued Pontryagin square terms
\beq 2\pi\int_\cM \frac{p_\tts}{4}\cP(b_\tts) + \frac{p_\ttc}{4}\cP(b_\ttc)~, \eeq
there is also a $\bZ_2$-valued cross term\footnote{
This corresponds to the universal quadratic group $U(\bZ_2\times\bZ_2)$ being $\bZ_4\times\bZ_4\times\bZ_2$.}
\beq 2\pi\int_\cM \frac{p_{\tts\ttc}}{2}b_\tts\cup b_\ttc~. \eeq
The quadratic function $\sigma$ \eqref{eqn.tildesigma} for the $PSO(2N)_{q_{ij},s_i}$ theory is
\begin{equation}
    \sigma(m_1,m_2) =  q_{ij}\frac{m_im_j}{4} + s_i\frac{m_i}{2} = \frac{1}{4}\left(q_{ij}+2\d_{ij}s_{j}\right)m_im_j \pmod{\mathbb Z}~, \label{quadratic function D series}
\end{equation}
and so $PSO(2N)_{q_{ij},s_i}$ theory can be obtained from $PSO(2N)_{0,0}$ by adding the discrete theta term
\begin{equation}
    2\pi\int_\mathcal{M} \left( \frac{p_\tts}{4}\mathcal{P}(b_\tts) + \frac{p_\ttc}{4} \mathcal{P}(b_\ttc) + \frac{p_{\tts\ttc}}{2}b_\tts \cup b_\ttc \right)~.
\end{equation}
where $p_\tts=q_{11}+2s_1\pmod{4}$, $p_\ttc=q_{22}+2s_2\pmod{4}$ and $p_{\tts\ttc}=q_{12}\pmod{2}$.

As was seen in \eqref{T transformation}, a $2\pi$-shift of the usual theta angle \eqref{theta term} is equivalent to shifting the lines by the quadratic function
\begin{equation}
    \sigma'(m_1,m_2) = \frac12 m_im_j \tr\m_{\circ i}^\vee\m_{\circ j}^\vee \pmod{\mathbb Z}~.
\end{equation}
Comparing this to \eqref{quadratic function D series} yields
\beq \Delta p_\tts = 2\tr\mu_{\circ1}^\vee\mu_{\circ1}^\vee = \frac{1}{2}N, ~~~
\Delta p_\ttc = 2\tr\mu_{\circ2}^\vee\mu_{\circ2}^\vee = \frac{1}{2}N,~~~
\Delta p_{\tts\ttc} = 2\tr\mu_{\circ 1}^\vee\mu_{\circ 2}^\vee = \frac{N-2}{2}\label{eqn.deltapso2n} \eeq
modulo $4$ for $\Delta p_\tts$ and $\Delta p_\ttc$ and modulo $2$ for $\Delta p_{\tts\ttc}$. The values of $\tr\m_\circ^{\vee i}\m_\circ^{\vee j}$ were read off from the entries of the inverse Cartan matrix \eqref{inverse-cartan-matrix-of-Dn}. Therefore, the $T$-transformation identifies the theories
\begin{equation}
     PSO(2N)^{\theta+2\pi}_{q_{ij},s_i} =  PSO(2N)^\theta_{q_{ij}+{\Delta p}_{ij},s_i}~,
\end{equation}
with $\Delta p_{ij}$ given in \eqref{eqn.deltapso2n}. In other words, shifting $\theta\mapsto\theta+2\pi$ is equivalent to adding the discrete theta term
\begin{equation}
    2\pi \int_\mathcal{M} \left( \frac{N}{8}\mathcal{P}(b_\tts+b_\ttc) +\frac{1}{2}b_\tts \cup b_\ttc \right)~, \label{coupling.for.spin4n}
\end{equation}
reproducing the result of~\cite{Cordova:2019uob}.
\subsection{\texorpdfstring{E Series: $\mathfrak{e}_6$ and $\mathfrak{e}_7$}{E Series: E6 and E7}}
Of the exceptional Lie groups, only $E_6$ and $E_7$ have non-trivial centers, with $Z(E_6)=\bZ_3$ and $Z(E_7)=\bZ_2$.
\subsection*{$\mathfrak{e}_6$}
The center of $E_6$ is isomorphic to $\mathbb Z_3$, and there are two possible gauge groups with $\mathfrak{e}_6$ algebra, namely $E_6$ and $E_6/\bZ_3$. The Dynkin diagram of the Lie algebra is
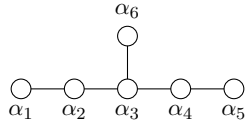
\begin{center}
\begin{dynkinDiagram}{E}{6}
\node [label={[scale=0.8,label distance=-0.8cm]:$\alpha_1$}] at (root 1) {};
\node [label={[scale=0.8]:$\alpha_6$}] at (root 2) {};
\node [label={[scale=0.8,label distance=-0.8cm]:$\alpha_2$}] at (root 3) {};
\node [label={[scale=0.8,label distance=-0.8cm]:$\alpha_3$}] at (root 4) {};
\node [label={[scale=0.8,label distance=-0.8cm]:$\alpha_4$}] at (root 5) {};
\node [label={[scale=0.8,label distance=-0.8cm]:$\alpha_5$}] at (root 6) {};
\end{dynkinDiagram}
\end{center}
Its center is generated by any of $\m^\vee_1$, $\m^\vee_2$, $\m^\vee_4$ or $\m^\vee_5$, and any of them furnishes a possible choice of basic magnetic charge. We choose $\m^\vee_\circ := \m^\vee_1$. The basic electric charge $\nu_\circ$ should be chosen so that $\moy{\nu_\circ,\mu^\vee_\circ}=1/3$. The quantization condition \eqref{DSZ non-abelian} has the following solutions, listed with the generating lines:
\begin{align}
    E_6&: (1,0)_\ttb, \quad (0,3)_\ttb~, \nonumber\\
    \left( E_6/{\mathbb{Z}_3} \right)_{q,s}&: (3,0)_\ttb, \quad (q,1)_s~,
\end{align}
where, for consistency, as discussed in \eqref{consistency conditions for SU(N)}, $q\equiv s \pmod{2}$. In the $E_6$ theory, $\theta$ is $2\pi$ periodic, while for the $E_6/{\mathbb{Z}_3}$, shifting $\theta\mapsto\theta+2\pi$ is equivalent to adding a discrete theta term of
\beq \Delta p = 3 C_{11}^{-1} = 4 \eeq
where, as in \eqref{q s and p}, $4p=q+3s\pmod{6}$. Thus, the following theories are identified
\begin{equation}
    \left( E_6/{\mathbb{Z}_3} \right)_{q,s}^{\theta+2\pi} = \left( E_6/{\mathbb{Z}_3} \right)_{q+4,s}^{\theta} = \left( E_6/{\mathbb{Z}_3} \right)_{q+1,s+1}^{\theta}~,
\end{equation}
and $\theta$ is $6\pi$ periodic. The three different $E_6/{\mathbb{Z}_3}$ theories are in the same $T$-orbit.
\subsection*{$\mathfrak{e}_7$}
The Dynkin diagram of $\mathfrak{e}_7$ is
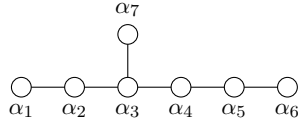
\begin{center}
\begin{dynkinDiagram}{E}{7}
\node [label={[scale=0.8,label distance=-0.8cm]:$\alpha_1$}] at (root 1) {};
\node [label={[scale=0.8]:$\alpha_7$}] at (root 2) {};
\node [label={[scale=0.8,label distance=-0.8cm]:$\alpha_2$}] at (root 3) {};
\node [label={[scale=0.8,label distance=-0.8cm]:$\alpha_3$}] at (root 4) {};
\node [label={[scale=0.8,label distance=-0.8cm]:$\alpha_4$}] at (root 5) {};
\node [label={[scale=0.8,label distance=-0.8cm]:$\alpha_5$}] at (root 6) {};
\node [label={[scale=0.8,label distance=-0.8cm]:$\alpha_6$}] at (root 7) {};
\end{dynkinDiagram}
\end{center}
and the center of the universal covering group $E_7$ is isomoprphic to $\mathbb{Z}_2$. It is generated by either $\m^\vee_4$, $\m^\vee_6$ or $\m^\vee_7$, and we choose $\m^\vee_\circ := \m^\vee_6$. Once again, the basic electric weight $\nu_\circ$ is chosen so that $\moy{\nu_\circ,\mu_\circ^\vee}=1/2$. Solving for the quantization conditions yields the following theories and corresponding generating lines
\begin{align}
    \leftidx{_s}{E_7}&: (1,0)_s, \quad (0,2)_\ttb~, \nonumber\\
    \left( E_7/{\mathbb{Z}_2} \right)_{q,s}&: (2,0)_\ttb, \quad (q,1)_s~. \nn
\end{align}
For $E_7/{\mathbb{Z}_2}$ theory, the discrete theta parameter can be identified as $p = q + 2s \pmod{4}$, and the $T$-transformation relates theories differing by a discrete theta term with
\beq \Delta p = 2C_{66}^{-1}=3~, \eeq
leading to the identifications
\begin{equation}
    \left( E_7/{\mathbb{Z}_2} \right)_{q,s}^{\theta+2\pi} = \left( E_7/{\mathbb{Z}_2} \right)_{q+3,s}^{\theta} = \left( E_7/{\mathbb{Z}_2} \right)_{q+1,s+1}^{\theta}~.
\end{equation}
Thus, $\theta$ is $8\pi$ periodic for $E_7/{\mathbb{Z}_2}$ theories, and all such theories are in the same $T$-orbit.

\section{Coupling to Background Fields \label{sec:background fields}}
\subsection{Abelian Theories}
\subsubsection{Symmetries}
$U(1)$ gauge theory, with action \eqref{eqn.maxwellaction}, has one-form symmetry group $U(1)_\tte\times U(1)_\ttm$~\cite{Gaiotto:2014kfa}, generated by the Noether currents
\beq j_\tte = \frac{1}{e^2}\ast\! f - \frac{\theta}{4{\pi}^2}f \,,\quad \quad j_\ttm = \frac{1}{2\pi} f~, \eeq
whose conservations follow from the equation of motion and the Bianchi identity respectively. In the notation of section \ref{charge lattice}, the integer-valued charges on a two-dimensional cycle $\Sigma$ are
\beq n = \int_\Sigma j_\tte~, \quad \quad m = \int_\Sigma j_\ttm~, \label{noether.charges}\eeq
which are related to the electric and magnetic charges by \eqref{electromagneticcharges}. There is a mixed 't Hooft anomaly between $U(1)_\tte$ and $U(1)_\ttm$, which can be explicitly seen in the next section when the theory is coupled to background gauge fields for the symmetries.

\subsubsection{Coupling to background gauge fields}
The background gauge field $B_\ttm\in H^2(\cM,U(1)_\ttm)$ for the magnetic symmetry $U(1)_\ttm$ couples to the action \eqref{eqn.maxwellaction} as
\beq S[B_\ttm] = \int_\cM -\frac{1}{2e^2}f\wedge\ast f+\frac{\theta}{2}\frac{f}{2\pi}\wedge\frac{f}{2\pi} + 2\pi\frac{B_\ttm}{2\pi}\wedge\frac{f}{2\pi}~. \label{eqn.maxwellbm} \eeq
Meanwhile, the coupling to the electric gauge field $B_\tte\in H^2(\cM,U(1)_\tte)$ is more complicated. To motivate it, we look at a simpler example of a two dimensional sigma model on target space $S^1$
\beq \int_\Sigma \frac{R^2}{2}\rd\phi \wedge\ast \rd\phi~, \eeq
which has an ordinary (zero-form) symmetry $U(1)_\ttt\times U(1)_\ttw$ with associated currents $j_\ttt=R^2\ast\rd\phi$ and $j_\ttw=\rd\phi/2\pi$. The translation symmetry $U(1)_\ttt$ is the analog of the electric symmetry of Maxwell theory. A background field $A_\ttt$ is a flat connection for some circle bundle over $\Sigma$. In order to couple the theory to $A_\ttt$, $\phi$ is promoted to a section of a bundle associated to said circle bundle, and the derivative is replaced with the covariant derivative
\beq \int_\Sigma \frac{R^2}{2} (\rd\phi+A_\ttt)\wedge\ast(\rd\phi+A_\ttt)~. \eeq
Analogously, in Maxwell theory, in order to couple to $B_\tte$, the gauge field $a$ is promoted to a bundle-valued section of a flat $U(1)$ gerbe, on which $B_\tte$ is a gerbe connection. The action is
\beq \int_\cM -\frac{1}{2e^2}(f+B_\tte)\wedge\ast(f+B_\tte)+\frac{\theta}{2}\frac{f+B_\tte}{2\pi}\wedge\frac{f+B_\tte}{2\pi}~. \label{eqn.maxwellbe} \eeq

Note that Maxwell theory cannot be simultaneously coupled to both $B_\tte$ and $B_\ttm$. Indeed, upon coupling to $B_\ttm$, the conservation of the current $j_\tte$ is violated by terms involving $B_\ttm$. Equivalently, if we took the action \eqref{eqn.maxwellbe} coupled to $B_\tte$ and tried to add the minimal coupling $B_\ttm\wedge f/2\pi$, we find that $B_\tte$ background gauge invariance is lost. It can be restored by introducing a bulk term on a five dimensional manifold $\cY$ with boundary $\p\cY=\cM$, as follows
\beq \int_\cM -\frac{1}{2e^2}(f+B_\tte)\wedge\ast(f+B_\tte)+ \frac{\theta}{2}\frac{f+B_\tte}{2\pi}\wedge\frac{f+B_\tte}{2\pi} + 2\pi\frac{B_\ttm}{2\pi}\wedge\(\frac{f}{2\pi}+\frac{B_\tte}{2\pi}\) - \int_\cY 2\pi \frac{B_\ttm}{2\pi}\wedge\frac{\rd B_\tte}{2\pi}~. \eeq
In other words, there is a mixed anomaly between the one form symmetries $U(1)_\tte$ and $U(1)_\ttm$, with anomaly polynomial
\beq 2\pi \int_\cY \frac{B_\ttm}{2\pi}\wedge\frac{\rd B_\tte}{2\pi}~., \label{eqn.maxwellanom} \eeq
which can be obtained via descent from the six dimensional polynomial~\cite{Gaiotto:2014kfa, Hsieh:2019iba}
\beq I_6 = \frac{\rd B_\ttm}{2\pi}\wedge\frac{\rd B_\tte}{2\pi}~.\eeq

\subsubsection{Relation to non-spin theories}
The non-spin theories we considered in section \ref{sec:abelian} can be obtained by turning on specific background fields. More specifically, we would like to couple a $\bZ_2$ subgroup of either the electric or magnetic $U(1)$ one-form symmetry to the second Stiefel-Whitney class $w_2(T\cM)$ of the spacetime manifold.

In the magnetic case, setting $B_\ttm=2\pi\frac{1}{2}w_2(T\cM)$ yields
\beq \int_cM -\frac{1}{2e^2}f\wedge\ast f+\frac{\theta}{2}\frac{f}{2\pi}\wedge\frac{f}{2\pi} + 2\pi\frac{1}{2}w_2(T\cM)c_1(E)~, \eeq
where $c_1(E)$ is the (mod $2$ reduction of the) first Chern class of the gauge bundle $E$. This discrete theta term makes the monopole a fermion~\cite{Wang:2018qoy}.

In the electric case, to gauge a $\bZ_2$ subgroup of $U(1)_\tte$, it is only necessary to promote the $U(1)$ bundle to a $U(1)/\bZ_2\simeq U(1)$ bundle, rather than a full blown gerbe. The quotient bundle has a characteristic class arising from the exact sequence
\beq 0\to \bZ_2 \to U(1) \to U(1)/\bZ_2 \to 0 \eeq
which we set to be equal to $B_\tte$. If $B_\tte=2\pi\frac{1}{2}w_2(T\cM)$, then the connection on the quotient bundle is exactly a $\Spin^\bC$ connection, which we saw in section \ref{sec:abelian} describes a theory with a fermionic electric particle.

If we turn on \emph{both} background gauge fields $B_\tte=B_\ttm=2\pi\frac{1}{2}w_2(T\cM)$, then we have a theory describing all-fermion electrodynamics. According to \eqref{eqn.maxwellanom}, this theory has an anomaly of~\cite{Wang:2013zja,Thorngren:2014pza,Hsin:2019fhf}
\beq 2\pi \int_\cY \frac{1}{2}w_2\beta(w_2) = 2\pi \int_\cY \frac{1}{2}w_2w_3~. \eeq
Here, $\beta$ is the Bockstein homomorphism associated to the sequence $0\to\bZ_2\to\bZ_4\to\bZ_2\to0$.\footnote{$\beta$ is also known as the first Steenrod square $\Sq^1$.}

\subsubsection{\texorpdfstring{$S$-duality}{S-duality}}
$S$-duality in Maxwell theory is an equivalence of theories under the exchange of electric and magnetic charges. In the path integral framework, it can be implemented by coupling the theory to a gauge field for its one-form electric symmetry, and integrating over the gauge field while inserting a Lagrange multiplier field coupled to the gauge field. Integrating out the Lagrange multiplier enforces flatness and triviality of the gauge field, returning us to the original Maxwell theory, while integrating out the gauge field instead leads one to the $S$-dual theory, formulated in terms of the Lagrange multiplier field~\cite{Witten:1995gf}.

For the non-spin Maxwell theories defined in section \ref{sec:abelian}, the theory with fermionic electric particle is $S$-dual to the theory with fermionic magnetic particle. This can be seen in a completely analogous manner, and was done in great detail in~\cite{Metlitski:2015yqa}. In the following, we give a brief outline of the procedure.

Let us review this procedure for spin spacetime manifolds. The partition function of Maxwell theory is
\beq Z[\tau] = \frac{1}{\vol(\cG)}\sum_{[f]\in H^2(\cM,\bZ)} \int [\rd a]\ \exp\(i S[a;\tau]\)~, \label{eqn.zmaxwell} \eeq
where the sum is over $U(1)$ bundles on $\cM$, and the path integral taken with respect to the connections on each bundle. The factor $\vol(\cG)$ is the volume of the group of gauge transformations of $a$. The Maxwell action is
\beq \int_\cM -\frac{1}{2e^2}f\wedge\ast f + \frac{\theta}{2}\frac{f}{2\pi}\wedge\frac{f}{2\pi} = \int_\cM \frac{i\bar\tau}{4\pi}f_+\wedge f_+ + \frac{i\tau}{4\pi}f_-\wedge f_-~, \label{eqn.smaxwell} \eeq
where we define $f_\pm:=\frac{1}{2}(f\pm\ast f)$ and recall \eqref{tau}
\beq \tau = \frac{\theta}{2\pi}+i\frac{2\pi}{e^2}~. \eeq
Now, we couple Maxwell theory to the electric background field $B_\tte$, and sum over $B_\tte$ while introducing a $U(1)$ connection-valued Lagrange multiplier $\tilde a$, with curvature $\tilde f$
\beq \int_\cM 2\pi \frac{B_\tte}{2\pi}\wedge\frac{\tilde f}{2\pi} = \int_\cM \frac{1}{2\pi}\(B_{\tte,+}\wedge\tilde f_+ + B_{\tte,-}\wedge\tilde f_-\) \eeq
yielding the partition function
\beq Z = \frac{1}{\vol(\cG)\vol(\tilde\cG)\vol(\cG_b)}\sum_{[f],[\tilde f]\in H^2(\cM,\bZ)} \int [\rd a\, \rd\tilde a\, \rd b_\tte]\ \exp\(i S[a,\tilde a,b_\tte;\tau]\)~, \eeq 
\begin{multline} S[a,\tilde a,b_\tte;\tau] = \int_\cM \frac{i\bar\tau}{4\pi}(f_++b_{\tte,+})\wedge(f_++b_{\tte,+}) + \frac{i\tau}{4\pi}(f_-+b_{\tte,-})\wedge(f_-+b_{\tte,-}) \\
+ \frac{i}{2\pi}b_{\tte,+}\wedge\tilde f_+ + \frac{i}{2\pi} b_{\tte,-}\wedge\tilde f_-~.
\end{multline}
If one integrates out $\tilde a$ first, the Lagrange multiplier term enforces the flatness of the gerbe connection $b_\tte$, and the sum over bundles on which $\tilde a$ is a connection enforces the topological triviality of $b_\tte$. This shows that the partition function is identical to that of Maxwell theory \eqref{eqn.zmaxwell}.

If instead one integrates out $b_\tte$ first, one can use its gauge symmetry to set $f=0$. The squares can be completed to
\begin{multline}
    S[a,\tilde a,b_\tte;\tau] = \int_\cM \frac{i\bar\tau}{4\pi}\(b_{e,+}+\frac{1}{\bar\tau}\tilde f_+\)\wedge\(b_{e,+}+\frac{1}{\bar\tau}\tilde f_+\) + \frac{i\tau}{4\pi}\(b_{e,-}+\frac{1}{\tau}\tilde f_-\)\wedge\(b_{e,-}+\frac{1}{\tau}\tilde f_-\) \\
    - \frac{i}{4\pi\bar\tau}\tilde f_+\wedge\tilde f_+ - \frac{i}{4\pi\tau}\tilde f_-\wedge \tilde f_-~.
\label{eqn.parentmaxwell}
\end{multline}
The gaussian integral over $b_\tte$ yields an overall $\tau$-dependent numerical factor,\footnote{
The factor is $\tau^{-(\chi+\sigma)/4}\bar\tau^{-(\chi-\sigma)/4}$~\cite{Witten:1995gf}.}
while the terms on the second line give exactly Maxwell theory, with connection $\tilde a$ and coupling constant $-1/\tau$.

On non-spin manifolds, there is a possibility of coupling Maxwell theory (electrically or magnetically) coupling to $w_2(T\cM)$. Consider first adding a magnetic coupling to \eqref{eqn.smaxwell} of the form
\beq 2\pi \int_\cM \frac{1}{2}w_2(T\cM)c_1(E) \eeq
where $c_1(E)$ is (the mod $2$ reduction of) the first Chern class of the line bundle. Upon coupling to $b_\tte$ and gauge fixing, this results in a five dimensional term
\beq 2\pi \int_\cY \frac{1}{2}w_2(T\cM)\frac{\rd b_\tte}{2\pi}, \eeq
and integrating out $b_\tte$ enforces the constraint
\beq \frac{\tilde f}{2\pi} = \frac{1}{2}w_2(T\cM) \pmod{\bZ}~. \eeq
In other words, the connection valued Lagrange multiplier $\tilde a$ is a $\Spin^\bC$ connection.

\subsection{Non-Abelian Theories}
\subsubsection{Symmetries}
Consider, as in section \ref{sec:non-abelian}, Yang-Mills theory based on a simple, connected and compact gauge group $G$. Let $\cL$ be its set of line operators, which, as we saw in section \ref{sec:Line Operators}, can be described as an extension of groups
\beq 0\to\hat\Gamma_\tte\to\cL\to\hat\Gamma_\ttm\to 0~, \label{eqn.lineext} \eeq
where $G=\tilde G/\Gamma_\ttm$ and $\Gamma_\tte=Z(G)$. The one-form symmetry group $\hat\cL$ is the Pontryagin dual $\hat\cL$
\beq 0\to\Gamma_\ttm\to\hat\cL\to\Gamma_\tte\to 0~. \label{eqn.groupext} \eeq
The charge operator -- the analogs of the Noether charges in the abelian case -- for the $\Gamma_\ttm$ subgroup is
\beq Q_\ttm(\Sigma) = \oint_\Sigma b_\ttm~, \eeq
where $b_\ttm\in H^2(\cM,\hat\Gamma_\ttm)$ is the Brauer class of the $G$-bundle, as we defined in section \ref{sec:Line Operators}. The unitary operator implementing the symmetry transformation by the group element $g \in \Gamma_\ttm$, is given by the exponential of the charge operator as
\beq U_\ttm(\Sigma,g)=e^{2\pi i \,\langle g, Q_\ttm\rangle}~. \eeq
The symmetry operators for the group elements which project non-trivially onto $\Gamma_\tte$ are disorder-type Gukov-Witten surface operators~\cite{Gukov:2006jk}. For ease of exposition, we denote the charge and unitary symmetry operator by
\beq Q(\Sigma)=\oint_\Sigma b~, \quad \quad U(\Sigma,g)=\exp\left(2\pi i \, g \oint_\Sigma b \right)~, \eeq
with $g \in \hat{\cL}$, keeping in mind that $b\in H^2(\cM,\cL)$ has a non-local expression in terms of fundamental fields of the theory. Let $\tilde b_\ttm\in C^2(\cM,\cL)$ be a lift of $b_\ttm$, and let $b_\tte\in C^2(\cM,\hat\Gamma_\tte)$ be defined by
\beq b = \tilde b_\ttm + b_\tte~. \label{symmetry.generators} \eeq
The conservation of $b$ leads to the relation
\beq \delta b_\tte = \hat\beta(b_\ttm)~, \eeq
where $\hat\beta:H^\ast(\cM,\hat\Gamma_\ttm)\to H^{\ast+1}(\cM,\hat\Gamma_\tte)$ is the Bockstein homomorphism\footnote{
$\hat\beta:H^p(--,\hat\Gamma_\ttm)\to H^{p+1}(--,\hat\Gamma_\tte)$ can also thought of as the extension class $\hat e\in H^{p+1}(B^p\hat\Gamma_\ttm,\hat\Gamma_\tte)$ of \eqref{eqn.lineext} as $p$-groups, under the relation $\hat\beta(b_\ttm)=b_\ttm^\ast\hat e$.}
associated to the extension \eqref{eqn.lineext}.

\subsubsection{Coupling to background gauge fields}
Let $A\in H^2(\cM,\hat\cL)$ be the background field for the $\hat\cL$ one-form symmetry. Let $A_\tte$ denote the projection of $A$ to $H^2(\cM,\Gamma_\tte)$, and $A_\ttm\in C^2(\cM,\Gamma_\tte)$ be defined by
\beq A = \tilde A_\tte + A_\ttm~, \eeq
where $\tilde A_\tte$ is some choice of lift of $A_\tte$. The flatness of $A$ implies
\beq \delta A_\ttm = \beta(A_\tte)~, \eeq
where $\beta:H^\ast(\cM,\Gamma_\tte)\to H^{\ast+1}(\cM,\Gamma_\ttm)$ is the Bockstein homomorphism\footnote{
Similarly, $\beta$ can be thought of as the extension class $e\in H^3(B^2\Gamma_\tte,\Gamma_\ttm)$ of \eqref{eqn.groupext} as higher groups, under the relation $\beta(A_\tte)=A_\tte^\ast e$.}
associated to the extension \eqref{eqn.groupext}. Physically, this reflects the fact that the symmetry group $\hat\cL$ is a $\Gamma_\ttm$-projective extension of $\Gamma_\tte$ -- so lifts of $\Gamma_\tte$ elements violate the group law up to elements of $\Gamma_\ttm$. This means that $\Gamma_\ttm$ flux can be sourced by junctions between $\Gamma_\tte$ symmetry operators~\cite{Tachikawa:2017gyf}.

In terms of the fields $A_\ttm$ and $A_\tte$, the minimal coupling can be written as
\beq 2\pi \int_\cM A \cup b = 2\pi\int_\cM A_\tte\cup b_\tte + A_\ttm\cup b_\ttm + \tilde A_\tte \cup \tilde b_\ttm~. \eeq
(Note that $A_\ttm\cup B_\tte=0$ due to exactness.)

\subsubsection{Anomalies} \label{sec.anomalies}
As in abelian gauge theory, there may exist mixed 't Hooft anomalies in the $\hat\cL$ one-form symmetry, which can be detected by coupling to the background gauge field $A$. Mixed anomalies between $\Gamma_\tte$ and $\Gamma_\ttm$ are characterized by the mild violation (i.e. violation only when background fields are turned on) of current conservation
\beq \delta b_\ttm = A_\tte^\ast\hat e~, \label{eqn.mixedanom} \eeq
where $\hat e\in H^3(B^2\Gamma_\tte,\hat\Gamma_\ttm)$ satisfies the condition
\beq e\cup\hat e = 0 \in H^6(B^2\Gamma_\tte,\bR/\bZ)~. \eeq
Fixing cocycle representatives for $e$, $\hat e$ and a representative $\omega\in C^5(B^2\Gamma_\tte,\bR/\bZ)$ satisfying $\delta\omega=e\cup\hat e$, the five-dimensional anomaly polynomial can be written as~\cite{Tachikawa:2017gyf}
\beq 2\pi \int_\cY A_\ttm \cup A_\tte^\ast \hat e - A_\tte^\ast\omega~. \label{eqn.thooftanom} \eeq

As an example of a theory exhibiting such an anomaly, consider the gauge group $SU(rs)/\bZ_s$, with $\eta=0$. The one-form symmetry of the theory is the direct product of $\Gamma_\tte=\bZ_r$ and $\Gamma_\ttm=\bZ_s$. The current for the magnetic symmetry is the Brauer class $b_\ttm=w_\ttm(E)\in H^2(\cM,\hat\Gamma_\ttm)$ of the gauge bundle $E$ associated to the sequence
\beq 0\to \hat\Gamma_\ttm=\bZ_s \to SU(rs) \to SU(rs)/\bZ_s \to 0~. \eeq
To couple to the background field $A_\tte$ for $\Gamma_\tte=\bZ_r$, one relaxes the $SU(rs)/\bZ_s$ gauge bundle to a $PSU(rs)$ gauge bundle, and sums over $PSU(rs)$ bundles in the path integral with Brauer class $w_\tte(E)\in H^2(\cM,\Gamma_\tte)$ associated to the sequence
\beq 0\to \Gamma_\tte=\bZ_r \to SU(rs)/\bZ_s \to PSU(rs) \to 0 \eeq
equal to the background field $A_\tte$. In other words, the action of the theory coupled to $A_\tte$ is
\beq S_{PSU(rs)}[a] + 2\pi \int_\cM \frac{1}{r}\tilde b_\tte\cup(w_\tte(E)-A_\tte)~, \eeq
where $S_{PSU(rs)}[a]$ is the Yang-Mills action for a $PSU(rs)$ gauge connection $a$, and $\tilde b_\tte$ is a $\bZ_r$-valued two-form Lagrange multiplier field enforcing $w_\tte(E)=A_\tte$.

$PSU(rs)$ bundles have a Brauer class $w\in H^2(\cM,\bZ_{rs})$ associated to
\beq 0\to \bZ_{rs} \to SU(rs) \to PSU(rs) \to 0 \eeq
whose mod $s$ reduction is the class mentioned above
\beq w(E) = w_\tte(E) \pmod{s}. \eeq
Meanwhile, the mod $s$-valued class $b_\ttm=w_\ttm(E)$ defined above for $SU(rs)/\bZ_s$-bundles are no longer well-defined here. Instead, we consider
\beq b_\ttm' = \frac{1}{r}\(\tilde w_\tte(E) - w(E)\)~, \label{eqn.bmprime} \eeq
where $\tilde w_\tte(E)\in C^2(\cM,\bZ_{rs})$ is some lift of $w_\tte(E)$. Note that the expression in brackets is an integer multiple of $r$, so \eqref{eqn.bmprime} defines a $\bZ_s$-valued cochain. The failure of the cocycle condition
\beq \delta b_\ttm' = \frac{1}{r}\delta\tilde w_\tte(E) = \beta(w_\tte(E)) = A_\tte^\ast\tilde e~, \eeq
is exactly given by the Bockstein homomorphism $\beta:H^2(\cM,\Gamma_\tte)\to H^3(\cM,\hat\Gamma_\ttm)$ acting on $w_\tte$, which can equivalently be thought of as the pullback of the extension class $\tilde e\in H^3(B^2\Gamma_\tte,\hat\Gamma_\ttm)$ of the one-form extension
\beq 0\to \hat\Gamma_\ttm \to Z(SU(rs)) \to \Gamma_\tte \to 0~. \eeq
Comparing this to \eqref{eqn.mixedanom}, we see that Yang-Mills theory with gauge group $SU(rs)/\bZ_s$ and $\eta=0$ has a mixed 't Hooft anomaly
\beq 2\pi\int_\cY A_\ttm \cup A_\tte^\ast\tilde e~. \eeq

Like in the abelian case, the spacetime manifold can couple to ($\bZ_2$ subgroups of) the one-form symmetries. For example, if $r$ is even, one could set $A_\tte= r\, w_2(T\cM)/2$, which makes the fundamental electric particles fermionic. If $s$ is even, setting $A_\ttm= s\, w_2(T\cM)/2$ makes the fundamental magnetic particles fermionic. In both $r$ and $s$ are even, turning on both couplings to $w_2(T\cM)$ results in the same all-fermion anomaly
\beq 2\pi\int_\cY \frac{1}{2}w_2\cup \Sq^1 w_2 = 2\pi\int_\cY \frac{1}{2}w_2w_3~, \eeq
as can be seen from \eqref{eqn.thooftanom}.

\subsubsection{\texorpdfstring{Anomalies and $T$-transformation}{Anomalies and T-transformation}} \label{sec.anomttrans}
Returning to the general setup at the beginning of section \ref{sec.anomalies}, consider now the addition of a discrete theta term
\beq 2\pi \int_\cM \cP_\sigma(b_\ttm)~. \eeq
Due to the anomaly \eqref{eqn.mixedanom}, the discrete theta term is not gauge invariant. Indeed,
\beq \delta S = 2\pi \int_\cY A_\tte^\ast e \cup \eta(b_\ttm) = 2\pi \int_\cY \eta(A_\tte^\ast e,\cup b_\ttm)~. \eeq
Comparing this with \eqref{eqn.thooftanom}, observe that this gauge variation can be cancelled by shifting the extension class $e$ by
\beq \Delta e = -\eta\circ\hat e~. \eeq
In other words, the addition of a discrete theta term modifies the extension class $e$ of the one-form symmetry group by an amount proportional to the 't Hooft anomaly $\hat e$.

\subsubsection{Gauging one-form symmetries} \label{gauging}
The various different gauge theories with based on the same Lie algebra can be related to one another by gauging their one-form symmetries. Indeed, we have already encountered and implicitly used this fact starting from the theory based on the simply connected Lie group $\tilde G$. The one-form symmetry group in this case is purely electric and equal to its center $Z(\tilde G)$. Suppose we minimally couple and dynamically gauge a subgroup $\hat\Gamma_\ttm\subset Z(\tilde G)$.\footnote{
Note that this is subtly different from coupling to a $Z(\tilde G)$ background field and then gauging a subgroup. There are some $\tilde G/\hat\Gamma_\ttm$ theories that cannot be obtained in this manner, whereas they can be reached by coupling directly to $\hat\Gamma_\ttm$ gauge fields.}
According to the prescription discussed above, this is done by promoting the $\tilde G$ bundle to a $\tilde G/\hat\Gamma_\ttm$ bundle, allowing cocycle conditions to be violated by elements in $\hat\Gamma_\tte$. The Brauer class $w(E)\in H^2(\cM,\hat\Gamma_\ttm)$ of the $\tilde G/\hat\Gamma_\ttm$ bundle is constrained to be equal to the gauge field $a_\ttm$ by a Lagrange multiplier field, which is then summed over in the path integral. Clearly, this simply yields $\tilde G/\hat\Gamma_\ttm$ theory with $\eta=0$, which now has an emergent $\Gamma_\ttm$ one-form symmetry generated by the current $w(E)$.

In order to arrive at the theories with non-trivial $\eta$, or, equivalently, non-trivial extensions
\beq 0\to \hat\Gamma_\tte \to \cL \to \hat\Gamma_\ttm \to 0~, \eeq
one couples non-minimally to the gauge field $a_\ttm$ by adding discrete theta terms of the form
\beq 2\pi \int_\cM \cP_\sigma(a_\ttm)~, \eeq
where $\sigma$ is the quadratic form corresponding to the desired theory, as discussed in section~\ref{sec:non-abelian}.

We can apply the same procedure to any theory, dynamically gauge some subgroup of its one-form symmetry $\hat\cL$, possibly with discrete theta terms, to arrive at any other theory with the same gauge algebra.

\subsubsection{Anomalies and one-form symmetries under gauging} \label{anomalies.and.symmetries}
Gauging a subgroup of the one-form symmetry yields another theory, with the same gauge algebra, but with the 't Hooft anomaly and group extension class exchanged. This is a manifestation of the phenomenon described in~\cite{Tachikawa:2017gyf}.

We illustrate this by beginning with a $\tilde G$ gauge theory, and gauge a subgroup $\Gamma_\ttm$ of its one-form symmetry group $Z(\tilde G)$ yielding $\tilde G/\Gamma_\ttm$ gauge theory. The parent $\tilde G$ theory has one-form symmetry group described by the extension \eqref{extension 2}. Let $e\in H^3(B^2\Gamma_\ttm,\Gamma_\tte)$ denote its extension class. This theory is non-anomalous, so the one-form symmetry of the daughter $\tilde G/\Gamma_\ttm$ theory is a direct product $\Gamma_\tte\times\hat\Gamma_\ttm$. Meanwhile, the background gauge field $b_\ttm$ of the parent theory is now summed over in the daughter theory, so the failure of the cocycle condition
\beq \delta\hat b_\ttm = A_\tte^\ast e \eeq
is now interpreted as a mixed anomaly with anomaly polynomial~\cite{Tachikawa:2017gyf}
\beq 2\pi \int_\cY A_\tte^\ast e\cup \hat A_\ttm~, \eeq
where $A_\tte,\hat A_\ttm$ are the two-form gauge fields for $\Gamma_\tte$ and $\hat\Gamma_\ttm$ respectively.

In order to arrive at other theories with gauge group $\tilde G/\Gamma_\ttm$ -- whose set of line operators $\cL$ is not a direct product of electric and magnetic factors, but rather a nontrivial extension \eqref{extension 1} -- we add a discrete theta term \eqref{eqn.distheta}. As discussed in section \ref{sec.anomttrans}, such a term shifts the extension class of the one-form symmetry by the mixed anomaly
\beq e' = \eta\circ e~, \eeq
where $\eta(x,y)=\sigma(x+y)-\sigma(x)-\sigma(y)$ is the bilinear form corresponding to the quadratic form used to define the Pontryagin square term in \eqref{eqn.distheta}. 

This discussion of the one-form symmetry gives an alternative understanding of \eqref{pullback} in terms of its Pontryagin dual. Indeed, $e\in H^3(B^2\Gamma_\tte,\Gamma_\ttm)$ is the extension class of \eqref{extension 2}, while $e'\in H^3(B^2\Gamma_\tte,\hat\Gamma_\ttm)$ is the extension class of the Pontryagin dual of \eqref{extension 1}
\beq 0\to \hat\Gamma_\ttm \to \hat\cL \to \Gamma_\tte \to 0~. \eeq
The relation $e'=\eta\circ e$ establishes the one-form symmetry group $\hat\cL$ as the pushout
\begin{equation} \label{pushout} \begin{tikzcd}
    0 \ar[r] &\hat\Gamma_\ttm \ar[r] &\hat\cL \ar[r] &\Gamma_\tte \ar[r] \ar[d,equal] &0 \\
    0 \ar[r] &\Gamma_\ttm \ar[r] \ar[u,"\eta"] &Z(\tilde G) \ar[r] \ar[u,dotted] &\Gamma_\tte \ar[r] &0
\end{tikzcd} \end{equation}
which is dual to the pullback diagram \eqref{pullback}.

\section*{Acknowledgments}
We would like to thank Zohar Komargodski, Martin Roček and Luigi Tizzano for their crucial insights and discussions. We especially thank Zohar Komargodski for suggesting this project. JPA and KR are partially supported by the NSF Grants PHY-1915093 and PHY-1620628. SS is supported in part by the Simons Foundation grant 488657 (Simons Collaboration on the Non-Perturbative Bootstrap).
\appendix

\section{Lie Algebra Basis and Weight Lattices} \label{app:lie.algebra}

We begin with a simple Lie algebra $\fg$, and choose a Cartan subalgebra $\ft\subset\fg$. Let $\alpha_i\in\ft^\ast$ denote the simple roots of $\fg$. The Cartan matrix is defined by
\begin{equation}
C_{ij} = 2 \frac{\tr\a_i\a_j}{\tr\a_j\a_j}~,~~~~~ 1\leq i,j\leq r
\end{equation} 
where $\tr$ is the Killing form on $\fg$ and $r$ is the rank of $\fg$. This does not depend on the overall normalization of the Killing form, but following an often used convention, we normalize it such that the length-squared of the long root(s) is $2$. The simple coroots $\alpha^\vee_i\in\ft$ are defined by
\beq \langle \a_i , \a^\vee_j \rangle =  2 \frac{\tr\a_i\a_j}{\tr\a_j\a_j}~, \eeq
which also does not depend on the overall normalization of the Killing form. The length-squared of the short coroot(s) is $2$. If we identify $\ft$ with $\ft^\ast$ using the Killing form, we get
\begin{equation}
    {\a_i^\vee}^\ast=\frac{\tr\a^\vee_i\a^\vee_i}{2} \a_i~, \quad\quad \text{and} \quad\quad (\tr\a^\vee_i\a^\vee_j)(\tr\a_i\a_i) = 4~.
\end{equation}
The corresponding fundamental weights are
\begin{equation}
\n_i= C^{-1}_{ij} \a_j~, \label{FW}
\end{equation}
which generate the weight lattice $\L_w$ of $\fg$. Note that although $C_{ij}$ has integral entries, its inverse generally does not. An irreducible representation of $\fg$ is characterized by its highest weight
\begin{equation}
\n= m_i \n_i, \quad m_i \in \mathbb{Z}^r~.
\end{equation}
Simple coroots $\alpha^\vee_i$ and fundamental coweights $\m^\vee_i$ of $\fg$ are defined as the dual vectors to the fundamental weights and simple roots. They generate respectively the coroot and coweight lattices $\L_{\ttcr}$ and $\L_{\ttcw}$. For convenience, we list here the pairings between fundamental weights, simple roots, fundamental coweights and simple coroots.
\begin{align}
\langle \a_i , \a^\vee_j \rangle &= C_{ij} ~, \\
\langle \n_i , \a^\vee_j \rangle &= \d_{ij}~,  \\
\langle \a_i , \m^\vee_j \rangle &= \d_{ij}~, \\
\langle \n_i , \m^\vee_j \rangle &= C^{-1}_{ij}~. \label{m dot m dual}
\end{align}
The coroots and fundamental coweights generate $\Lambda_\ttcr $ and $\Lambda_\ttcw$ respectively and we have
\begin{align}
    \Lambda_\ttr &\subset \Lambda_{G} \subset \Lambda_\ttw \subset \ft^\ast~,\\
    \Lambda_\ttcr  &\subset \Lambda_{\mathrm{c}G} \subset \Lambda_\ttcw \subset \ft~.
\end{align}
Here $\Lambda_{\mathrm{c}G}$ is the cocharacter lattice of $G$ where the magnetic charges take value in, and its dual lattice $\Lambda_{G}:=\Hom(\Lambda_{\mathrm{c}G},\mathbb{Z})$ is the character lattice of $G$. The lattices $\Lambda_{G}$ and $\Lambda_{\mathrm{c}G}$ are dual in the sense that for any $\n \in \Lambda_{G}$ and $\m^\vee \in \Lambda_{\mathrm{c}G}$, $\langle \n,\m^\vee \rangle\in\mathbb Z$. As explained in section \ref{sec:one-form symmetries}, the actual weight lattice of a theory with gauge group $G$ is an extension of $\Lambda_{\mathrm{c}G}$ by $\Lambda_{G}$, where the extension is determined by the topological theta parameters of the theory. In particular modding out by the (co)root lattices and projecting onto the one-form symmetry charges, we get the extension \eqref{extension 1}. For weights $\nu\in\Lambda_\ttw$ and $\mu^\vee\in\Lambda_\ttcw$, we denote the corresponding one-form symmetry charges by $[\nu]\in\Lambda_\ttw/\Lambda_\ttr$ and $[\mu^\vee]\in\Lambda_\ttcw/\Lambda_\ttcr$, which are the projection of the weights onto $\Lambda_\ttw/\Lambda_\ttr$ and $\Lambda_\ttcw/\Lambda_\ttcr$.

\section{Adjoint Higgsing} \label{app.coulomb}

In this section we perform a consistency condition of our proposal by adjoint Higgsing. We add an adjoint scalar Higgs field to the non-abelian theories we discussed. As a result, the theory flows in the infrared to an abelian gauge theory with gauge group $\mathbb{T}=\ft/\Lambda_{\ttc G}$--which is the maximal (Cartan) torus of $G$--with some topological theta term. We refer to this infrared theory as the Coulomb phase. This abelian theory is free and as discussed in section \ref{sec:abelian} we know its set of line operators and their spins exactly. We see how our proposal predict the spins of these lines in the infrared.

Denoting the gauge field of the Higgsed theory as $A \in \Omega^1(\cM,\ft)$, the gauge transformation are given by
\begin{equation}
    A \mapsto A + \rd \lambda~, \quad \mathrm{for:} \quad \lambda \in \Omega^0(\cM,\mathbb{T})~,
\end{equation}
where $\lambda:\cM \to \mathbb{T}$ is a smooth function. Every such map can be written as
\begin{equation}
    \lambda = \exp(i \lambda^i h_i^\vee )~,
\end{equation}
where $h_1^\vee,\dots,h_{r=\mathrm{rank}(G)}^\vee$ generate the cocharacter lattice of $G$,
\begin{equation}
    \Lambda_{\ttc G} = \{m_1 h_1^\vee + \cdots + m_r h_r^\vee \mid m_i \in \bZ \}~.
\end{equation}
This lattice is the kernel of the exponential map above and we have the short exact sequence
\begin{equation}
    0 \to \Lambda_{\ttc G} \to \ft \xrightarrow{\exp} \mathbb{T} \to 0~.
\end{equation}
Therefore $\lambda^i : \cM \to \bR/\bZ$, and each of them individually define a separate $U(1)$ gauge transformation. Accordingly if we expand the gauge fields as
\begin{equation}
    A = A^i h_i^\vee~, \quad \mathrm{where:} \quad A^i \in \Omega^1(\cM,\bR)~,
\end{equation}
each of $A^i$ is a seperate $U(1)$ gauge field with $U(1)$ holonomies. Upon Higgsing, the theory flows in the infrared to
\begin{equation}
S =\int_{\mathcal{M}_4} \left(-\frac{1}{2 e^2}g_{ij} F^i\wedge * F^j + \frac{\theta}{8 \pi^2}g_{ij} F^i\wedge F^j\right)~,
\end{equation}
where $g_{ij}=\tr h_i^\vee h_j^\vee$, and $F^i = \rd A^i$.

Denote $Q_{i\tte}$ and $Q_{i\ttm}$ as the electric and magnetic charges under $A^i$. More precisely
\begin{equation}
    Q_{i\tte} = \frac{1}{e} \oint_\Sigma \ast F^i~, \quad \quad  Q_{i\ttm} = \frac{1}{e} \oint_\Sigma F^i~,
\end{equation}
measure the electric and magnetic charges inside the surface $\Sigma$. By doing the Noether procedure as in \eqref{noether.charges}, we find the integer valued one-form symmetry charges
\begin{equation}
    n_{i} = g_{ij} \oint_\Sigma \left(\frac{1}{e^2} \ast F^j - \frac{\theta}{4\pi^2} F^j\right)~, \quad \quad  m_{i} = \frac{1}{2\pi} \oint_\Sigma F^i~,
\end{equation}
where we have the relations
\begin{equation}
    n_i = g_{ij} \left(\frac{1}{e}Q_{j\tte}-\frac{e\theta}{4\pi^2}Q_{j\ttm}\right)~, \quad\quad m_i = \frac{e}{2\pi} Q_{i\ttm}~. \label{emcharges.inhiggspahse}
\end{equation}
Therefore, it is easy too see that we have
\begin{equation}
    Q_{i\tte} \, g^{-1}_{ij} \,Q'_{j\ttm} - Q'_{i\tte} \, g^{-1}_{ij} \, Q_{j\ttm} \in 2\pi \bZ~, \label{DSZ.Higgs}
\end{equation}
for any pair of dyons. This is in fact the DSZ quantization condition of this theory, which also measures the angular momentum stored in the electromagnetic fields in the presence of the pair of dyons. Now one could find the Wilson-'t Hooft lines which are charged under these one-form symmetries and match them with the UV line operators. We define the Wilson lines by integrating the gauge fields, and the 't Hooft lines can be defined as the boundary of the surfaces which generate the electric one-form symmetries,
\begin{equation}
    W^i(\gamma) = \exp\left(2\pi i \oint_\gamma A^i \right)~, \quad\quad T^i(\partial \Sigma) = \exp\left(2\pi i \,g_{ij}\int_\Sigma \left(\frac{1}{e^2} \ast F^j - \frac{\theta}{4\pi^2} F^j\right) \right)~.
\end{equation}
We find the UV line operators that flows after Higgsing to these basic Wilson and 't Hooft lines, by finding their corresponding weights. First we pick a basis for the character lattice $\Lambda_G$ of $G$ which is dual to the chosen basis for $\Lambda_{\ttc G}$, i.e.
\begin{equation}
    \langle h_i, h^\vee_j \rangle = \d_{ij}~.
\end{equation}
Thus in the UV weights of the form $(h_i,h^\vee_j)$ generate all the possible genuine Wilson-'t Hooft lines. Wilson lines of the UV theory are labeled by representations of $G$. Define $R_i$ as the representation whose highest weight is $h_i$, then the corresponding UV Wilson line is defined by
\begin{equation}
    W(R_i) = \tr_{R_i} \exp\left(i \oint A\right) = 
    \sum_{\nu \in R_i} \exp\left(i \oint \langle \nu,A \rangle \right) = \exp\left(i \oint  A^i \right) + \cdots~,
\end{equation}
where the sum is over all the weights $\nu$ of the representation $R_i$. In particular, $h_i$ is such a weight and since $\langle h_i , A\rangle=A^i$, the sum includes the IR Wilson line $W^i$. Thus we see that the UV Wilson line, decomposes into a sum of IR Wilson lines which should all carry the same spin as the UV line. Thus having a labeling for the UV Wilson lines, we infer the spins of the IR lines. Similarly one could see that the UV 't Hooft line characterized by the coweight $h_i^\vee$, flows in the IR into a sum including basic IR 't Hooft line $T^i$. For these IR lines, the angular momentum of the electromagnetic fields of two dyons with charges $(n_i,m_i)$ and $(n_i',m_i')$ is given by the Dirac pairing \eqref{DSZ.Higgs} as
\begin{equation}
    2J_3 = \sum_{i=1}^{r} n_im_i' - n_i'm_i = \big\langle (n_ih_i,m_ih^\vee_i),(n_i'h_i,m_i'h^\vee_i) \big\rangle_\mathrm{D} \pmod{2}~.
\end{equation}
Thus the Dirac pairing of the UV lines labeled by weights $(n_ih_i,m_ih^\vee_i)$ and $(n_i'h_i,m_i'h^\vee_i)$, indeed calculate the angular momentum of the electromagnetic fields in this Coulomb phase. Thus our proposal of labeling the UV lines with spin, is consistent with the spin of the IR lines after Higgsing.

Also note that all the 't Hooft lines of the IR theory whose weights (charges) are inside the coroot lattice $\Lambda_\ttcr$, can be realized as dynamical 't Hooft-Polyakov monopole solutions of the UV theory. This is because, the VEV of the Higgs field breaks the guage group $G$ down to its Cartan torus $\mathbb{T}$. Thus topologically solitonic solutions are characterized by $\pi_2(G/\mathbb{T})$. There is the long exact sequence
\begin{equation}
    \pi_2(G)=1 \to \pi_2(G/\mathbb{T}) \to \pi_1(\mathbb{T})= \Lambda_{\ttc G} \to \pi_1(G) = \Lambda_{\ttc G}/\Lambda_{\ttcr} \to \cdots~,
\end{equation}
and we get $\pi_2(G/\mathbb{T})=\Lambda_\ttcr$. Thus 't Hooft lines with charges $(0,m_i)$ such that $m_ih^\vee_i \in \Lambda_\ttcr$, are UV completed as 't Hooft-Polyakov monopoles of the UV theory.

Furthermore, from \eqref{emcharges.inhiggspahse} one could find the effect of shifting $\theta$ by $2\pi$ by noting that
\begin{equation}
    n_i = g_{ij} \left(\frac{1}{e}Q_{j\tte}-\frac{\theta}{2\pi}m_j\right)~.
\end{equation}
Hence shifting $\theta$ by $2\pi$ shifts the electric weight $\nu=n_ih_i$ by
\begin{equation}
    \Delta \nu = -g_{ij}m_jh_i = -{(m_j h^\vee_j)}^\ast~,
\end{equation}
which is consistent with the $T$-transformation defined in section \ref{sec:rules}.
\section{S and T Transformations} 
\label{App:S and T nonabelian}
Here we study the $S$ and $T$ operations defined in \cite{Gaiotto:2014kfa} (section 6), on the non-abelian gauge theories discussed in the previous section. These operations are defined as a generalization of Witten's operation on 3d theories with global symmetries \cite{Witten:2003ya}. The $T$ operation is already discussed in section \ref{sec:T transformation}, and acts by adding a discrete theta term to the action. Whereas the $S$ operation acts by gauging the one-form symmetry of the theory which is further explained in section \ref{sec:background fields}. The $S$ and $T$ operations can be defined by their action on the line operators as
\begin{align}
    S&: (\n,\m^\vee)_s \mapsto ({\m^\vee}^\ast,-{\n}^\ast)_s~,\\
    T&: (\n,\m^\vee)_s \mapsto (\n - {\m^\vee}^\ast,\m^\vee)_s~,
\end{align}
where the map $^\ast:\ft \to \ft^\ast$ is induced by the Killing form on $\ft$--see section \ref{sec:T transformation}. Since different gauge theories are determined by their set of genuine line operators and their spins, these maps uniquely determine a map between the theories. It is straightforward then to obtain the full $SL(2,\mathbb{Z})$ orbits of these non-abelian gauge theories. We do it for $\mathfrak{u}(2)$ and $\mathfrak{su}(4)$ gauge theories below.

Note that these operations, a priori, do not act on the coupling constant $\tau$ of the theory. But in some cases if we also change $\tau$ we can obtain duality transformations. For instance as was explained in section \ref{sec:T transformation}, for the $T$ operation by shifting the $\theta$-angle of the theory by $2\pi$, the $T$ operation becomes a duality transformation. Furthermore, for the case of $U(1)$ theories by a proper action on $\tau$ we have the full $SL(2,\mathbb{Z})$ duality transformation on the abelian theories. However, for non-abelian theories to get the full $SL(2,\mathbb{Z})$ group to act as a duality transformation, we need to look at the supersymmetric version of the Yang-Mills theory.

\subsection*{$\mathcal{N}=4$ Super Yang-Mills on Non-Spin Manifolds}
For the case of $\mathcal{N}=4$ super Yang-Mills theory, the $S$ and $T$ operations become duality transformations \cite{Montonen:1977sn,Osborn:1979tq,Vafa:1994tf} given the actions
\begin{equation}
	S : \tau \mapsto -\frac{1}{n_{\mathfrak{g}}\tau}~, \quad T : \tau \mapsto \tau+1~,
\end{equation}
on the coupling constant of the theory, where $n_{\mathfrak{g}}=1$ for simply laced Lie algebras (for non-simply laced algebras see \cite{Argyres:2006qr, Kapustin:2006pk}). The orbits of $S$-duality for non-abelian theories on spin manifolds was obtained in \cite{Aharony:2013hda}. Here we extend that discussion to non-spin manifolds.

As explained in \cite{Cordova:2018acb}, to define $\mathcal{N}=2$ super Yang-Mills theory with $SU(2)$ guage group on non-spin manifolds, we have to define a non-abelian spin/charge relation \cite{Freed:2016rqq, Wang:2018qoy} because of the adjoint fermions. In particular, we have to identify the $\mathbb{Z}_2$ subgroup of the Lorentz transformation generated by $(-1)^F$ with the central $\mathbb{Z}_2$ subgroup of the global $R$-symmetry that the adjoint fermions are charged under. For the case of $\mathcal{N}=4$ supersymmetry, the adjoint fermions transform in the fundamental representation of the $SU(4)_R$ symmetry, so one could identify the central element $-\mathbbm{1} \in SU(4)_R$ with $(-1)^F$ of the Lorentz group. This defines a non-abelian spin/charge relation which allows us to define the $\mathcal{N}=4$ Yang-Mills theory on non-spin manifolds.
\subsection{\texorpdfstring{$\fg=\mathfrak{su}(2)$}{g=su(2)}}
For the example of $\mathfrak{su}(2)$ gauge theory on non-spin manifolds, all the theories belong to the same orbit of $S$ and $T$ operations and we get the following web
\begin{equation}\label{su(2) orbit}
\begin{tikzcd}
    &&SO(3)_{1,\ttf} \arrow[rd, rightarrow,"T"] \arrow[dd, leftrightarrow,"S"]&&\\
	SU(2) \arrow[loop, leftrightarrow,"T"] \arrow[r,leftrightarrow,"S"] & SO(3)_{0,\ttb} \arrow[ru, rightarrow,"T"] && SO(3)_{0,\ttf} \arrow[dl, rightarrow,"T"] \arrow[r,leftrightarrow,"S"] & \text{Spin-}SU(2) \arrow[loop, leftrightarrow,"T"]\\
	&&SO(3)_{1,\ttb} \arrow[lu, rightarrow,"T"]&&
\end{tikzcd}
\end{equation}
By looking at the $\mathcal{N}=4$ version of these theories, we can obtain extra consistency checks for our proposal by Higgsing these theories to $U(1)$, and match the $S$-duality orbits.

For the case of $\leftidx{_{s'}}{SU(2)}{}$ theory, the fundamental Wilson line has spin $s'$ which under Higgsing becomes the fundamantal Wilson line of the $U(1)$ theory. For the $SO(3)_{q,s}$ theory however, the basic Wilson line is $W^2=(2,0)_\ttb$ which Higgses to the fundamental Wilson line of the $U(1)$ theory $(1,0)_\ttb$. The other generating line $W^qT=(q,1)_s$, Higgses to a line of spin $s$ and electric charge $q/2$ in the $U(1)$ theory. This fractional electric charge is equivalent to adding a theta angle of $q\pi$ to the $U(1)$ theory. Moreover by looking at the coupling constants of these theories, we have the relations $\tau^{U(1)}=2\tau^{SU(2)}$ and $\tau^{U(1)}=\tau^{SO(3)}/2$. Therefore after Higgsing we get
\begin{equation}
    \begin{tikzcd}
        \leftidx{_{s'}}{SU(2)}{^\tau}: (1,0)_{s'}, (0,2)_\ttb \ar[d,"\text{Higgsing}"] & {SO(3)}_{q,s}^{\tau} : (2,0)_\ttb, (q,1)_s \ar[d,"\text{Higgsing}"] \\
        \leftidx{_{s'}}{{U(1)}}{_{\ttb}^{2\tau}}: (1,0)_{s'}, (0,1)_\ttb & {U(1)}_{s}^{(\tau+q)/2}: (1,0)_\ttb,(0,1)_s
    \end{tikzcd}
\end{equation}
To see the consistency of the $S$-duality of $\mathfrak{su}(2)$ theories with $U(1)$, we Higgs the theories in \eqref{su(2) orbit} and indeed we get a web of dualities for the $U(1)$ theories as expected
\begin{equation}
\begin{tikzcd}
    &&U(1)_{\ttf}^{(2\tau-1)/2\tau} \arrow[rd, equal]&&\\
	U(1)_\ttb^{2\tau} \arrow[r,leftrightarrow,"S"] & U(1)_{\ttb}^{-1/2\tau} \arrow[ru, rightarrow,"T"] && U(1)_{\ttf}^{(2\tau-1)/2\tau} \arrow[dl, rightarrow,"T"] \arrow[r,leftrightarrow,"S"] & \left(\text{Spin}^\mathbb{C}\right)_{\ttb}^{2\tau/(1-2\tau)}\\
	&&U(1)_{\ttb}^{(4\tau-1)/2\tau} \arrow[lu, leftarrow,"T^2"]&&
\end{tikzcd}
\end{equation}
%
\subsection{\texorpdfstring{$\fg=\mathfrak{su}(4)$}{g=su(4)}}
For the case of Yang-Mills theory based on the Lie algebra $\mathfrak{su}(4)$, we get the following orbits
\begin{equation*}
{\scalefont{0.77}
\begin{tikzcd}[column sep=1.6em, row sep=1.1em]
    \tikzstyle{every node}=[font=\tiny]
    &&&PSU(4)_{2,\ttb}\arrow[rd, rightarrow,"T"]\arrow[dd, leftrightarrow,"S"]&&&\\
    &&PSU(4)_{1,\ttf}\arrow[ru, rightarrow,"T"]\arrow[dddd, leftrightarrow,"S"]&&PSU(4)_{3,\ttf}\arrow[rdd, rightarrow,"T"]\arrow[dddd, leftrightarrow,"S"]&&\\
    &&&SO(6)_{1,\ttf}\arrow[dd, leftrightarrow,"T"]&&&\\
    SU(4)\arrow[swap,loop,leftrightarrow,"T"]\arrow[r,leftrightarrow,"S"]&PSU(4)_{0,\ttb}\arrow[ruu, rightarrow,"T"]&&&&PSU(4)_{0,\ttf}\arrow[ddl, rightarrow,"T"]\arrow[r, leftrightarrow,"S"]&\text{Spin-}SU(4)\arrow[swap,loop,leftrightarrow,"T"]\\
    &&&SO(6)_{1,\ttb}\arrow[dd, leftrightarrow,"S"]&&&\\
    &&PSU(4)_{3,\ttb}\arrow[uul, rightarrow,"T"]&&PSU(4)_{1,\ttb}\arrow[dl, rightarrow,"T"]&&\\
    &&&PSU(4)_{2,\ttf}\arrow[lu, rightarrow,"T"]&&&
\end{tikzcd}
}
\end{equation*}
\begin{equation*}
    {\scalefont{0.78}
    \begin{tikzcd}
        SO(6)_{0,\ttb} \arrow[loop left, leftrightarrow,"S"] \arrow[r, leftrightarrow,"T"] & SO(6)_{0,\ttf} \arrow[r, leftrightarrow,"S"] & {\text{Spin-}SO(6)}_\ttb \arrow[loop right, leftrightarrow,"T"] &&& {\text{Spin-}SO(6)}_\ttf \arrow[loop right, leftrightarrow,"S\text{, }T"]
    \end{tikzcd}}.
\end{equation*}
Interestingly for this example, different orbits have different one-form symmetry groups or the same symmetry but different 't Hooft anomalies for that symmetry. In particular, the theories in the orbit of $SU(4)$ have a $\mathbb{Z}_4$ one-form symmetry, whereas the other theories have a $\mathbb{Z}_2 \times \mathbb{Z}_2$ one-form symmetry. Furthermore, as explained in section \ref{sec:background fields}, the theories in the orbit of ${\text{Spin-}SO(6)}_\ttb$ are anomaly free while the ${\text{Spin-}SO(6)}_\ttf$ theory has a gravitational 't Hooft anomaly.

\section{Consistency with Wu's Formula}
\label{App: Wu}
In this Appendix we perform a non-trivial check on our proposal of determining the spin of lines in non-Ableian gauge theories. This check is based on Wu's formula, which states that for any $\bZ_2$ cohomology class $x \in H^2(\cM,\bZ_2)$,
\beq  \frac12 w_2(T\cM) \cup x + \frac12 x \cup x \equiv 0 \pmod{2}~, \label{wu.formula} \eeq
where $w_2(T\cM)$ is the second Stiefel-Whitney class of (the tangent bundle of) the manifold $\cM$. In the examples of section \ref{sec:non-abelian}, take $x=b_\ttm$ as a generator of some $\bZ_2$ one-form symmetry. Adding ($2\pi$ times) the LHS of \eqref{wu.formula} to the action, we get a duality since the action remains invariant mod $2\pi$. This duality puts some constraints on the set of line operators and their spin. To see this, note that the first term in \eqref{wu.formula} attaches a $w_2(T\cM)$ surface to those lines which are charged under this $\bZ_2$ symmetry, and hence changes their spin. Whereas the second term attaches the surface
\beq \exp\left( 2\pi i \int_\Sigma x \right)~, \label{surface} \eeq
to the charged line operators supported on $\gamma=\partial \Sigma$. But such open surfaces are not gauge invariant, and has to be attached to a line operator with some weight which we denote by $\Delta \tilde{\g} \in \tilde{\Lambda}$. Therefore, if we start with a theory that have a line operator with weight $\tilde{\gamma}\in \tilde{\cL}$ and spin $s \pmod{2}$ which is charged under the $\bZ_2$ symmetry, then the duality guarantees that the theory should also have a line with spin $s+1 \pmod{2}$ and weight $\tilde{\gamma} + \Delta \tilde{\g}$. Now we want to prove this requirement by using our proposal.

Let us assume that the theory has a set of lines labeled by their weights in $\tilde{\cL}$. Thus the one-form symmetry charges of these lines is given by projection of their weights onto $\cL$. This projection is given by the restriction of the projection $p:\tilde{\Lambda} \to \Lambda$ to $\tilde{\cL} \subset \tilde{\Lambda}$. Also, since $\cL$ is the group of the charges, its Pontryaging dual $\hat{\cL}$, is the symmetry group. Denote $b\in H^2(\cM, \cL)$ as the generator of this one-form symmetry, and denote the $\bZ_2$ one-form symmetry subgroup by the inclusion $i:\bZ/2\bZ \to \hat{\cL}$. The Pontryagin dual of this inclusion defines the generator of this $\bZ_2$ one-form by
\beq  \hat{i}(b) =: x \in H^2(\cM,\bZ/2\bZ) ~,  \eeq
where we have identified $\bZ_2$ and $\hat{\bZ}_2$ with $\bZ/2\bZ$. Now take an arbitrary line operator of the theory with weight $\tilde{\gamma}$ and one-form charge $\g = p(\tilde{\g})$, then the $\bZ_2$ charge of this line is given simply by $\hat{i}(\g) \in \bZ/2\bZ$. We conclude that adding the first term in \eqref{wu.formula} to the action change the spin of a line with weight $\tilde{\g}$ by $\hat{i}(\g)$. So it changes the spin lables $s:\tilde{\cL} \to \hat{\bZ}_2$ to $s'=s+\hat{i}\circ p$. Now we want to find how the weights change by adding the second term in \eqref{wu.formula}, i.e. finding $\Delta\tilde{\g}$. To do so note that the second coupling in \eqref{wu.formula} can be written as
\beq \eta(b,\cup b) = \frac12 \hat{i}(b) \cup \hat{i}(b) \pmod{\bZ}~, \label{def.of.eta} \eeq
where $\eta:\cL\times\cL \to \bR/\bZ$. Equivalently, $\eta$ can be viewed as a homomorphism $\eta:\cL\to\Lambda/\cL$, where we have identified $\widehat{\cL}$ with $\Lambda/\cL$ by an isomorphism\footnote{More precisely the Dirac pairing on $\Lambda$--which is the projection of the pairing \eqref{pairing} on $\Lambda$--gives an isomorphism ${}^\ast:\hat{\cL} \to \Lambda/\cL$, by the relation
\begin{equation}
    \langle g^\ast, \g \rangle_{\mathrm{Dirac}} = \langle g, \g \rangle \pmod{\bZ} \quad \mathrm{for:} \quad g\in\hat{\cL}~~\mathrm{and}~~\g\in\cL~,
\end{equation}
where the RHS is evaluated by the natural pairing between $\widehat{\cL}$ and $\cL$. This can be summarized in a short exact sequence $0 \to \cL \to \Lambda \to \hat{\cL} \to 0$. Interestingly when the sequence is not split, the $\hat{\cL}$ one-form symmetry is anomalous.} induced by the Dirac pairing on $\Lambda$. Then similar to \eqref{the.lattice}, $\eta$ changes the Lagrangian charge lattice $\cL\subset\Lambda$ to\footnote{However for $\eta$ given in \eqref{def.of.eta} $\cL'=\cL$, otherwise the duality would have failed.}
\begin{equation}
    {\cL}' = \{\g+\Delta\g \mid \g\in\cL\,,~\Delta\g \equiv 2 \eta(\g) \mod{\cL}\}~,
\end{equation}
and we have $\Delta \tilde{\g} \in 2p^{-1}(\eta(\g))$, where here $p$ is the projection $\tilde{\Lambda}\to\Lambda/\cL$.

Putting everything together the duality sends the line with weight $\tilde{\g}$ with spin $s(\tilde{\g})$ to the line with weight $\tilde{\g} + \Delta \tilde{\g}$ and spin $(s+\hat{i}\circ p)(\tilde{\g})$ which has to be the genuine line of the same theory. But before the action of this duality the line with weight $\tilde{\g} + \Delta \tilde{\g}$ has spin $s(\tilde{\g} + \Delta \tilde{\g})$. Thus the consistency of our proposal with Wu's formula requires these two spins to be the same, i.e.
\begin{equation}
    \begin{split}
    (s+\hat{i}\circ p)(\tilde{\g}) &\equiv s(\tilde{\g} + \Delta \tilde{\g}) \pmod{2\bZ} \\
    s(\tilde{\g})+\hat{i}\circ p(\tilde{\g}) &\equiv s(\tilde{\g}) + s(\Delta \tilde{\g})+\langle \Delta \tilde{\g}, \tilde{\g} \rangle_{\mathrm{Dirac}} \pmod{2\bZ} \\
    \hat{i}(\g) &\equiv 2 \langle  \eta(\g), \g \rangle_{\mathrm{Dirac}} \pmod{2\bZ} \\
    \hat{i}(\g) \hat{i}(\g) &\equiv 2 \eta(\g,\g) \pmod{2\bZ}
\end{split}
\end{equation}
where in the second line we have used $s(\Delta \tilde{\g}) \equiv 0 \pmod{2\bZ}$, this is because $\Delta \tilde{\g}$ is inside $2p^{-1} (\eta(\g))$ and hence is twice of some weight, so according to \eqref{spin label} is bosonic. The last line simply follows from the definition of $\eta$ in \eqref{def.of.eta}. Thus we have shown the consistency of our proposal with Wu's formula.
\section{Normalization of theta terms} 
\label{App:NormTheta}
In this Appendix, we discuss our normalization of the continuous theta term
\beq \frac{\theta}{2}\int_\cM \tr \frac{f}{2\pi}\wedge\frac{f}{2\pi}~. \eeq
We shall do so by specifying the characteristic class, as well as any numerical factor, which multiplies the periodic parameter $\theta$. This will unambiguously give the definition of $\theta$, avoiding any niggly numerical factors, and furthermore, the integrality and multiplicity of characteristic classes will immediately tell us the periodicity of $\theta$.

For abelian $U(1)$ theory, the theta term is normalized as
\beq \frac{\theta}{2}\int_\cM c_1(E)^2~, \eeq
where $c_1(E)$ is the first Chern class of the $U(1)$ bundle $E$. On spin manifolds, $c_1^2$ is even, and so $\theta$ is $2\pi$-periodic for a spin theory. On non-spin manifolds, however, $\theta$ is $4\pi$-periodic.

For non-abelian theories, the story is more complicated, as there are different Lie groups corresponding to the same Lie algebra. The fourth cohomology of a connected, simply connected and simple Lie group $\tilde G$ has rank one
\beq H^4(B\tilde G,\bZ)=\bZ~, \eeq
and let $\tilde x$ be its generator (fixed up to sign of $\theta$). For example, in the case $\tilde G=SU(N)$, $\tilde x$ is the second Chern class $c_2$ of the $N$-dimensional complex bundle; for $\tilde G=Sp(N)$, $\tilde x$ is the first Pontrjagin class $p_1$ of the $2N$-(complex) dimensional quaternionic bundle. The theta term an oriented Yang-Mills theory with gauge group $\tilde G$ is normalized as
\beq \theta \int_\cM a^\ast\tilde x~, \eeq
where $a:\cM\to B\tilde G$ is the Yang-Mills connection. It follows that $\theta$ is $2\pi$ periodic.

Now, consider a group $G=\tilde G/\pi_1(G)$ which is not simply connected. The fourth integral group cohomology of $BG$ is again of rank one, and the projection map $\rho:\tilde G\to G$ induces an injection
\beq \rho^\ast:H^4(BG,\bZ)\to H^4(B\tilde G,\bZ)~. \eeq
Let $x$ be a generator of $H^4(BG,\bZ)$. Then
\beq \rho^\ast x=n\tilde x \eeq
for some nonzero integer $n\in\bZ$, which we can choose to be positive by redefining $x$. The theta term of an oriented Yang-Mills theory with gauge group $G$ is normalized as
\beq \frac{\theta}{n} \int_\cM a^\ast x~, \eeq
and it follows that $\theta$ is $2\pi n$ periodic.

For example, for $G=PSU(N)=SU(N)/\bZ_N$, $n=N$ if $N$ is odd and $n=2N$ if $N$ is even~\cite{Cordova:2019uob}. One way of computing $n$ is to compute $c_2(\operatorname{Ad})$ for the associated adjoint bundle $\operatorname{Ad}$, which is a characteristic class of $PSU(N)$, in terms of characteristic classes of the defining $N$-dimensional bundle. One finds that $c_2(\operatorname{Ad})=2Nc_2(N)$~\cite{Witten:2000nv}. For $N$ even, this class is primitive and is therefore equal to $\rho^\ast x$ (up to sign), while for $N$ odd, this class is divisible by $2$ and so $2\rho^\ast x=c_2(\operatorname{Ad})$ (up to sign)~\cite{Gu2016,Cordova:2019uob}.

\begin{figure}
\centering
\begin{tabular}{|c|c|c|c|}
\hline
$G$ & $\tilde G$ & $n$ & $\Delta p \pmod{n}$ \\ \hline
$PSU(N)$, $N$ even & $SU(N)$ & $2N$ & $N-1$ \\ \hline
$PSU(N)$, $N$ odd & $SU(N)$ & $N$ & $-1$ \\ \hline
$PSp(N)$ & $Sp(N)$ & $2$ & $N$ \\ \hline
$SO(N)$, $N\geq 5$ & $Spin(N)$ & $2$ & $1$ \\ \hline
$PSO(2N)$, $N$ odd & $Spin(2N)$ & $8$ & $N$ \\ \hline
$PSO(2N)$, $N$ even & $Spin(2N)$ & $8$ &  see \eqref{coupling.for.spin4n} \\ \hline
$E_6/\bZ_3$ & $E_6$ & $3$ & $1$ \\ \hline
$E_7/\bZ_2$ & $E_7$ & $2$ & $1$ \\ \hline
\end{tabular}
\caption{Table of gauge groups, $\theta$ periodicities and $T$-transformations}
\label{tbl.gndeltap}
\end{figure}

\section{Relation between discrete and continuous theta terms} 
\label{App:reltheta}
The relation between discrete and continuous theta terms can be understood as a relation between integral and torsion characteristic classes of the gauge $G$-bundle. See section 6 of~\cite{Aharony:2013hda} for a similar discussion on spin spacetime manifolds. As in Appendix \ref{App:NormTheta}, let $x$ be a generator of $H^4(BG,\bZ)$, and the positive integer $n$ be such that the continuous theta term is normalized as $\theta x/n$.

When $\theta$ is a multiple of $2\pi$, the continuous theta term can be expressed in terms of a discrete theta term $\cP(w)$, where $b$ is the Brauer class of the $G$ bundle corresponding to the extension $1\to\Gamma_\ttm\to\tilde G\to G\to 1$. This is the topological genesis of the $T$-transformation described in section \ref{sec:T transformation}. The relation takes the form
\beq x = -\Delta p\, \cP(b)\pmod{n}~. \eeq
The values of theta period $n$ and $T$-transformation $\Delta p$ are tabulated in table \ref{tbl.gndeltap}. For some gauge groups such as $PSU(N)$ and $SO(N)$, the relations can be found and corroborated in the mathematics literature -- see~\cite{10.2307/1993484,Gu2016}, as well as~\cite{Aharony:2013hda,Cordova:2019uob} and the references therein. For other gauge groups, we are not aware of results in the mathematics literature, and the physical arguments in this and the earlier papers~\cite{Aharony:2013hda,Cordova:2019uob} serve as heuristic arguments for them.

\bibliographystyle{JHEP}
\bibliography{refs}

\end{document}